\documentclass[a4paper, 12pt]{article}
\usepackage{float}
\usepackage{natbib}
\usepackage{graphicx}
\usepackage{xcolor}
\usepackage{multirow}
\usepackage{subfig}
\usepackage{amssymb}
\usepackage{amsmath}
\usepackage{amsthm}
\DeclareMathOperator{\trace}{tr}

\DeclareMathOperator{\BetaD}{Beta}

 \DeclareMathOperator{\ND}{N}

 \DeclareMathOperator{\IGD}{IG}

 \def \calD {\mathcal D}

\def \bvec {\text{\boldmath$b$}}    
    
    \def \mD {\text{\boldmath$D$}}

    \def \mI {\text{\boldmath$I$}}
    
    \def \mK {\text{\boldmath$K$}}

\def \uvec {\text{\boldmath$u$}}

\def \yvec {\text{\boldmath$y$}}    
    
\def \nuvec {\text{\boldmath$\nu$}}

\def \xtildevec {\text{\boldmath$\tilde x$}}

\def \betavec         {\text{\boldmath$\beta$}}

\def \thetavec        {\text{\boldmath$\theta$}}
\def \varthetavec     {\text{\boldmath$\vartheta$}}

\def \nuvec           {\text{\boldmath$\nu$}}

\def \varthetahatvec     {\text{\boldmath$\hat \vartheta$}}

\def \mSigma   {\mathbf{\Sigma}}

\def \nullvec {\mathbf{0}}
\def \onevec {\mathbf{1}}

\title{Bayesian Measurement Error Correction in Structured Additive Distributional Regression with an Application to the Analysis of Sensor Data on Soil-Plant Variability}
\author{Alessio Pollice, Giovanna Jona Lasinio,\\ Roberta Rossi, Mariana Amato,\\ Thomas Kneib, Stefan Lang}

\begin{document}
\maketitle
\begin{abstract}
The flexibility of the Bayesian approach to account for covariates with measurement error is combined with semiparametric regression models
for a class of continuous, discrete and mixed univariate response distributions with potentially all parameters depending on a structured
additive predictor. Markov chain Monte Carlo enables a modular and numerically efficient implementation of Bayesian measurement error
correction based on the imputation of unobserved error-free covariate values. We allow for very general measurement errors, including
correlated replicates with heterogeneous variances. The proposal is first assessed by a simulation trial, then it is applied to the
assessment of a soil-plant relationship crucial for implementing efficient agricultural management practices. Observations on multi-depth
soil information forage ground-cover for a seven hectares Alfalfa stand in South Italy were obtained using sensors with very refined
spatial resolution. Estimating a functional relation between ground-cover and soil with these data involves addressing issues linked to the
spatial and temporal misalignment and the large data size. We propose a preliminary spatial interpolation on a lattice covering the field
and subsequent analysis by a structured additive distributional regression model accounting for measurement error in the soil covariate.
Results are interpreted and commented in connection to possible Alfalfa management strategies.
\end{abstract}


\section{Introduction}

Standard regression theory assumes that explanatory variables are deterministic or error-free, but this assumption is quite unrealistic for
many biological processes and replicated observations of covariates are often obtained to quantify the variability induced by the presence
of measurement error (ME). Indeed covariates measured with error are considered a serious danger for inferences drawn from regression
models. The most well known effect of measurement error is the bias towards zero induced by additive i.i.d. measurement error, but under
more general measurement error specifications (as considered in this paper), different types of misspecification errors are to be expected
\citep{Carroll:2006, Loken584}. This is particularly true for semiparametric additive models, where the functional shape of the relation
between responses and covariates is specified adaptively and therefore is also more prone to disturbances induced by ME. Recent papers
advocate the hierarchical Bayesian modeling approach as a natural route for accommodating ME uncertainty in regression models.
%
%
%
%
In particular, in the context of semiparametric additive models, \cite{sarkar2014} provide a Bayesian model based on B-spline mixtures and
Dirichlet process mixtures relaxing some assumptions about the ME model, such as normality and homoscedasticity of the underlying true
covariate values. Indeed many authors rely on almost flat normal priors for the true covariate, as \cite{muff2015} who frame ME adjustments
into Bayesian inference for latent Gaussian models with the integrated nested Laplace approximation (INLA). The INLA framework allows to
incorporate various types of random effects into generalized linear mixed models, such as independent or conditional auto-regressive models
to account for continuous or discrete spatial structures. Relatively few articles have explicitly addressed covariate ME in the context of
spatial modeling. A notable exception is given by the work of \cite{arima2017} who proposed a parametric spatial model for multivariate
Bayesian small area estimation, accounting for heteroscedastic ME in some of the covariates.
Continuous spatial data are approached by \cite{huque2016} to assess the relationship between a covariate with ME and a spatially
correlated outcome in a non-Bayesian semiparametric regression context, assuming that the true unobserved covariate can be modeled by a
smooth function of the spatial coordinates.

In this paper we introduce a functional ME modeling approach allowing for replicated covariates with ME within a flexible class of
regression models recently introduced by \cite{Klein:2015}, namely structured additive distributional regression models. In this modeling
framework, each parameter of a class of potentially complex response distributions is modeled by an additive composition of different types
of covariate effects, e.g. non-linear effects of continuous covariates, random effects, spatial effects or interaction effects. Structured
additive distributional regression models are therefore in between the simplistic framework of exponential family mean regression and
distribution-free approaches as quantile regression. We allow for quite general measurement error specifications including multiple
replicates with heterogeneous dependence structure. From a computational point of view, based on the seminal work by \cite{berry2002} for
Gaussian scatterplot smoothing and \cite{Kneib2010} for general semiparametric exponential family and hazard regression models, we develop
a flexible fully Bayesian ME correction procedure based on Markov chain Monte Carlo (MCMC) techniques to generate observations from the
joint posterior distribution of structured additive distributional regression models. ME correction is obtained by the imputation of
unobserved error-free covariate values in an additional sampling step. Our implementation is based on an efficient binning strategy that
avoids recomputing the complete design matrix after imputing true covariate values and combines this with efficient storage and computation
schemes for sparse matrices.

The main motivation of our investigation comes from a case study on the use of proximal soil-crop sensor technologies to analyze the
within-field spatio-temporal variation of soil-plant relationships in view of the implementation of efficient agricultural management
practices. More precisely, we analyze the relationship between multi-depth soil information indirectly assessed through the use of high
resolution geophysical soil proximal sensing technology and data of forage ground-cover variation measured by a multispectral radiometer
within a seven hectares Alfalfa stand in South Italy. Observations of both quantities were made using sensors with very refined spatial
resolution: ground-cover data were obtained at four sampling occasions with point locations changing over time, while soil data were
sampled only once for three different depth layers. Estimating a functional relation between ground-cover and soil with the data at hand
involves addressing several issues, also linked to the spatial and temporal misalignment and the large data size. The nonlinear relation
between crop productivity and soil is estimated by additive distributional regression models with structured additive predictor and
measurement error correction. While distributional regression allows to deal with the heterogeneity of the response scale at the four
sampling occasions, the ME correction is motivated by observations of covariates being replicated along a depth gradient and extends the
model proposed by \cite{Kneib2010}, accounting for heterogeneous variances and possibly dependent replicates of the soil covariate.


The paper is structured as follows: in Section~\ref{sec:memodel}, we introduce the Bayesian additive distributional regression model and
the ME specification. Section~\ref{sec:inf} reports details of the MCMC estimation algorithm and different tools for model choice and
comparison. Simulation-based investigations of the model performance are the contents of Section~\ref{sec:sims}. Section~\ref{sec:case} is
dedicated to review some issues of the sampling design, describe the spatial interpolation method, report summary features of the data at
hand, interpret estimation results and comment on possible Alfalfa management strategies. The final Section~\ref{sec:summ} summarizes our
main findings and includes comments on directions for future research.

\section{Measurement Error Correction in Distributional Regression}\label{sec:memodel}

The main motivation for our methodological developments comes from the need to estimate the nonlinear dependence of ground-cover on soil information
by a smooth function, accounting for the heterogeneity in the position and scale of the response due to the sampling time, for the repeated
measurements of the soil covariate and for the residual variation of unobserved spatial features. A detailed discussion of the data
collection and pre-processing is given in Section~\ref{sec:case}. There we address ground-cover differences in both location and scale by
structured additive distributional regression models, in which both parameters of the response distribution are related to additive
regression predictors \citep[][and references therein]{Klein:2015}. The explicit consideration of replicated covariate observations
avoids erroneous conclusions about the precise functional form of the relationship between ground-cover and soil variables or even about
the presence of any association. In this section, the distributional regression model is set up, structured additive predictors and the
measurement error correction are introduced with the relevant prior distributions.

\subsection{Distributional Regression}\label{subs:model}

Our treatment of Bayesian measurement error correction is embedded into the general framework of structured additive distributional
regression. Assume that independent observations $(y_i,\nuvec_i)$ $i=1,\ldots,n$ are available on the response $y_i$ and covariates
$\nuvec_i$ and that the conditional distribution of the responses belongs to a $K$-parametric family of distributions such that
$y_i|\nuvec_i\sim\calD(\varthetavec(\nuvec_i))$ and the $K$-dimensional parameter vector $\varthetavec(\nuvec_i) =
(\vartheta_1(\nuvec_i),\allowbreak\ldots,\allowbreak\vartheta_K(\nuvec_i))'$ is determined based on the covariate vector $\nuvec_i$. More
specifically, we assume that each parameter is supplemented with a regression specification
\[
 \vartheta_k(\nuvec_i) = h_k(\eta^{\vartheta_k}(\nuvec_i))
\]
where $h_k$ is a response function that ensures restrictions on the parameter space and $\eta^{\vartheta_k}(\nuvec_i)$ is a regression predictor.

In our analyses, we will consider two specific special cases where $y_i\sim\ND(\mu(\nuvec_i), \allowbreak\sigma(\nuvec_i)^2)$, i.e. the responses are
conditionally normal with covariate-dependent mean and variance and $y_i\sim\BetaD(\mu(\nuvec_i), \sigma^2(\nuvec_i))$, i.e. conditionally
beta distributed responses with regression effects on location and scale. To ensure $\sigma(\nuvec_i)^2>0$ for the variance of the normal distribution, we use the exponential link
function, i.e. $\sigma(\nuvec_i)^2 = \exp(\eta^{\sigma^2}(\nuvec_i))$ while for both parameters $\mu(\nuvec_i)$ and $\sigma(\nuvec_i)^2$ of the beta distribution we employ a logit
link since they are restricted to the unit interval \citep{FerCri2004}.

\subsection{Structured Additive Predictor}\label{subs:pred}

For each of the predictors, we assume an additive decomposition as
\[
 \eta^{\vartheta_k}(\nuvec_i) = \beta_0^{\vartheta_k} + f_1^{\vartheta_k}(\nuvec_i) + \ldots + f_{J_k}^{\vartheta_k}(\nuvec_i)
\]
i.e. each predictor consists of a total of $J_k$ potentially nonlinear effects $f_j^{\vartheta_k}(\nuvec_i)$, $j=1,\ldots,J_k$, and an additional overall intercept $\beta_0^{\vartheta_k}$. The nonlinear effects $f_j^{\vartheta_k}(\nuvec_i)$ are a generic representation for a variety of different effect types (including nonlinear effects of continuous covariates, interaction surfaces, spatial effects, etc., see below for some more details on examples that are relevant in our application). Any of these effects can be approximated in terms of a linear combination of basis functions as
\[
 f(\nuvec_i) = \sum_{l=1}^L\beta_lB_l(\nuvec_i) = \bvec_i'\betavec
\]
where we dropped both the function index $j$ and the parameter index $\vartheta_k$ for simplicity, $B_l(\nuvec_i)$ denotes the different basis functions with basis coefficients $\beta_l$ and $\bvec_i=(B_1(\nuvec_i),\ldots,B_L(\nuvec_i))'$ and $\betavec=(\beta_1,\ldots,\beta_l)'$ denote the corresponding vectors of basis function evaluations and basis coefficients, respectively.

Since in many cases the number of basis functions will be large, we assign informative multivariate Gaussian priors
\[
 p(\betavec|\thetavec) \propto \exp\left(-\frac{1}{2}\betavec'\mK(\thetavec)\betavec\right)
\]
to the basis coefficients to enforce certain properties such as smoothness or shrinkage. The specific properties are determined based on the prior precision matrix $\mK(\thetavec)$ which itself depends on further hyperparameters $\thetavec$.

To make things more concrete, we discuss some special cases that we will also use later in our application:
\begin{itemize}
 \item Linear effects: for parametric, linear effects, the basis functions are only extracting certain covariate elements from the vector $\nuvec_i$ such that $B_l(\nuvec_i) = \nu_{il}$. In case of no prior information, flat priors are obtained by $\mK(\thetavec)=\nullvec$ which reduces the multivariate normal prior to a multivariate flat prior. In case of a rather large number of parametric effects, a Bayesian ridge prior can be obtained by $\mK(\thetavec) = \tau^2\mI_L$ where $\tau^2$ is an additional prior variance parameter that can, for example, be supplemented an inverse gamma prior.
 \item Penalised splines for nonlinear effects $f(x)$ of continuous covariates $x$: for nonlinear effects of continuous covariates, we follow the idea of Bayesian penalized splines \citep{BreLan2006} where B-spline basis functions are combined with a random walk prior on the regression coefficients. In this case, the prior precision matrix is of the form $\mK(\thetavec) = \tau^2\mD'\mD$ where again $\tau^2$ is a prior variance parameter and $\mD$ is a difference matrix. To obtain a data-driven amount of smoothness, we will assign an inverse gamma prior $\tau^2\sim\IGD(a,b)$ to $\tau^2$ with $a=b=0.001$ as a default choice.
 \item Tensor product splines for coordinate-based spatial effects $f(s_x, s_y)$: for modelling spatial surfaces based on coordinate information, we utilize tensor product penalized splines where each of the basis functions is constructed as $B_l(s_x,s_y) = B_{l,x}(s_x)B_{l,y}(s_y)$ with univariate B-spline bases $B_x(\cdot)$ and $B_y(\cdot)$. The penalty matrix is then given by
     \[
     \mK(\thetavec) = \frac{1}{\tau^2}\left[\omega\mK_x \otimes \mI_{y} + (1-\omega)\mI_{x}\otimes\mK_y\right]
     \]
     where $\mK_x$ is a random walk penalty matrix for a univariate spline in $s_x$, $\mK_y$ is a random walk penalty matrix for a univariate spline in $s_y$ and $\mI_x$ and $\mI_y$ are identity matrices of appropriate dimension. In this case, the prior precision matrix comprises two hyperparameters: $\tau^2$ is again a prior variance parameter (and can therefore still be assumed to follow an inverse gamma distribution) while $\omega\in(0,1)$ is an anisotropy parameter that allows for varying amounts of smoothness along the $x$ and $y$ coordinates of the spatial effect. For the latter, we assume a discrete prior following the approach discussed in \citet{KneKleLanUml2017}.
\end{itemize}
For some alternative model components comprise spatial effects based on regional data, random intercepts and random slopes or varying coefficient terms, see \citet[][Ch.~8]{Fahrmeir:2013}

\subsection{Measurement Error}\label{subsec:me}

In our application, we are facing the situation that a continuous covariate with potentially nonlinear effect modelled as a penalized spline has been observed with measurement error. Hence, we are interested in estimating the nonlinear effect $f(x)$ in one of the predictors of a distributional regression model where instead of the continuous covariate $x$ we observe $M$ replicates
\[
 \tilde{x}_i^{(m)} = x_i + u_i^{(m)}, \quad m=1,\ldots,M,
\]
contaminated with measurement error $u_i^{(m)}$. For the measurement error, we consider a multivariate Gaussian model such that
\[
 \uvec_i\sim\ND_M(\nullvec, \mSigma_{u,i})
\]
where $\uvec_i = (u_i^{(1)},\ldots,u_i^{(M)})'$ and $\mSigma_{u,i}$ is a known, pre-specified covariance matrix. Independent replicates
with constant variance were considered in \citet{Kneib2010} for a mean regression model specification, with
$\boldsymbol{\Sigma}_{u;i}=\sigma^2_{u}\mathbf{I}_M\label{eq:MEcov1}$. This assumption is here relaxed in the context of distributional
regression, considering correlated replicates with potentially heterogeneous variances and covariances, leaving $\boldsymbol{\Sigma}_{u;i}$
unstructured.
Of course, it would be conceptually straightforward to include priors on unknown parameters in $\mSigma_{u,i}$, but this would require
large amounts of data to obtain reliable estimates. We therefore restrict ourselves to the case of a known covariance matrix.

The basic idea in Bayesian measurement error correction is now to include the unknown, true covariate values $x_i$ as additional unknowns
to be imputed by MCMC simulations along with estimating the other parameters in the model. This requires that we assign a prior
distribution to $x_i$ as well and rely on the simplest version
\[
 x_i\sim\ND(\mu_x, \tau_x^2)
\]
where we achieve flexibility by adding a further level in the prior hierarchy via
\[
 \mu_x\sim\ND(0, \tau_\mu^2), \qquad \tau_x^2\sim\IGD(a_x, b_x).
\]
To obtain diffuse priors on these hyperparameters, we use $\tau_\mu^2=1000^2$ and $a_x=b_x=0.001$ as default settings.

\section{Bayesian Inference}\label{sec:inf}

All inferences rely on MCMC simulations implemented within the free, open source software BayesX \citep{Belitz:2013}. As described in
\citet{Lang:2014}, the software makes use of efficient storing even for large data sets and sparse matrix algorithms for sampling from
multivariate Gaussian proposal distributions. In the following, we mostly focus on the required sampling steps for the measurement error
part, since the remaining parts are mostly unchanged compared to standard distributional regression models \citep[see][for
details]{KleSohKneLan2015}.

\subsection{Measurement Error Correction}\label{subsec:infme}

For the general case of distributional regression models, no closed form full conditional is obtained for the true covariate values $x_i$
such that we rely on a Metropolis Hastings update. Proposals are generated based on a random walk proposal
\begin{equation}\label{eq:propo}
 x_i^p = x_i^c + e_i, \qquad e_i\sim\ND(0, f \cdot \trace(\mSigma_{u,i})/M^2)
\end{equation}
where $x_i^p$ denotes the proposal, $x_i^c$ is the current value and the variance of the random walk is determined by the measurement error
variability (quantified by the trace of the covariance matrix $\mSigma_{u,i}$), the number of replicates $M$ and a user-defined scaling
factor $f$ that can be used to determine appropriate acceptance rates. The proposed value is then accepted with probability
\[
 \alpha(x_i^p|x_i^c) = \frac{p(y_i|x_i^p)}{p(y_i|x_i^c)} \frac{\exp\left(-\frac{1}{2\tau_x^2}(x_i^p-\mu_x)^2\right)}{\exp\left(-\frac{1}{2\tau_x^2}(x_i^c-\mu_x)^2\right)}
 \frac{\exp\left(-\frac{1}{2}(\xtildevec_i-x_i^p\onevec_M)'\mSigma_{u,i}^{-1}(\xtildevec_i-x_i^p\onevec_M)\right)}{\exp\left(-\frac{1}{2}(\xtildevec_i-x_i^c\onevec_M)'\mSigma_{u,i}^{-1}(\xtildevec_i-x_i^c\onevec_M)\right)}
\]
where the first term is the ratio of likelihood contributions given the proposed and current values of the covariate, the second term is
the ratio of the measurement error priors and the third term is the ratio of the measurement error likelihoods. The ratio of the proposal
densities cancels due to the symmetry of the random walk proposal.

In contrast, the updates of the hyperparameters are standard due to the conjugacy between the Gaussian prior for the true covariate values
and the hyperprior specifications. We therefore obtain
\[
 \mu_x|\cdot\sim\ND\left(\frac{n\bar{x}\tau_\mu^2}{n\tau_{\mu}^2 + \tau_x^2},\frac{\tau_x^2\tau_\mu^2}{n\tau_{\mu}^2 + \tau_x^2}\right)
\]
and
\[
 \tau_x^2|\cdot\sim\IGD\left(a_x + \frac{n}{2}, b_x + \frac{1}{2}\sum_{i=1}^n(x_i-\mu_x)^2\right)
\]

As mentioned before, priors could be assigned to incorporate uncertainty about elements of the covariance matrices $\mSigma_{u;i}$ of the measurement error
model. However, reliable estimation of $\mSigma_{u;i}$ would typically require a large number of repeated measurements on the covariates,
therefore we will restrict our attention to the case of known measurement error covariance matrices $\mSigma_{u;i}$.

\subsection{Updating the Structured Addditive Predictor}\label{subsec:infpred}

Updating the components of a structured additive predictor basically follows the same steps as in any distributional regression model
\citep[see][for details]{KleSohKneLan2015,KneKleLanUml2017}. One additional difficulty arises from the fact that the imputation of true
covariate values in each iteration implies that the associated spline design matrix would have to be recomputed in each iteration which
would considerably slow down the MCMC algorithm. To avoid this, we utilize a binning approach \citep[see][]{LanUmlWecHarKne2012} where we
assign each exact covariate value to a small interval. Using a large number of intervals allows us to control for potential rounding
errors. On the positive side, the binning approach allows us to precompute the design matrix for all potential intervals and to only
re-assign the observations based on an index vector in each iteration instead of recomputing the exact design matrix.

\subsection{Model Evaluation}\label{subsec:eval}

We combine different tools to determine a final model considering the quality of estimation and the predictive ability in terms of
probabilistic forecasts. We rely on the deviance information criterion \citep[DIC,][]{Spiegelhalter:2002}, the Watanabe-Akaike information
criterion \citep[WAIC,][]{Watanabe:2010}, proper scoring rules \citep{Gneiting:2007} as well as normalized quantile residuals
\citep{Dunn:1996}.

Measures of predictive accuracy are generally referred to as \emph{information criteria} and are defined based on the deviance (the log
predictive density of the data given a point estimate of the fitted model, multiplied by -2). Both DIC and WAIC adjust the log predictive
density of the observed data by subtracting an approximate bias correction. However, while DIC conditions the log predictive density of the data
on the posterior mean $\mbox{E}(\thetavec|\yvec)$ of all model parameters $\thetavec$ given the data $\yvec$, WAIC averages it over the posterior distribution $p(\thetavec|\yvec)$. Then, compared to DIC, WAIC has
the desirable property of averaging over the whole posterior distribution rather than conditioning on a point estimate. This is especially
relevant in a predictive Bayesian context, as WAIC evaluates the predictions that are actually being used for new data while DIC estimates
the performance of the plug-in predictive density \citep{Gelman2014}.

As measures of the out-of-sample predictive accuracy we consider three \emph{proper scoring rules} based on the logarithmic score
$S(f_r,y_r)=\log(f_r(y_r))$, the spherical score $S(f_r,y_r)=f_r(y_r)/\|f_r(y_r)\|_2$ and the quadratic score
$S(f_r,y_r)=2f_r(y_r)-\|f_r(y_r)\|_2^2$, with $\|f_r(y_r)\|_2^2=\int f_r(\omega)^2 d\omega$. Here $y_r$ is an element of the hold out
sample $y_1,\ldots,y_R$ and $f_r$ is the predictive distribution of $y_r$ obtained in the current cross-validation fold.
In practice the data set is approximately divided into $R$ non-overlapping subsets of equal size and predictions for one of the subsets are
obtained from estimates based on all the remaining subsets. The predictive ability of the models is then compared by the aggregated average
score $S_R=(1/R) \sum_{r=1}^R S(f_r,y_r)$. Higher logarithmic, spherical and quadratic scores deliver better forecasts when comparing two
competing models.
With respect to the others, the logarithmic scoring rule is usually more susceptible to extreme observations that introduce large contributions in the log-likelihood \citep{Klein:2015}.

If $F$ and $y_i$ are respectively the assumed cumulative distribution and a realization of a continuous random variable and $\varthetahatvec$ is
an estimate of the distribution parameters, \emph{quantile residuals} are given by $\hat r_i=\Phi^{-1}(u_i)$, where $\Phi^{-1}$ is the
inverse cumulative distribution function of a standard normal distribution and $u_i=F(y_i|\varthetahatvec_i)$. If the estimated model is close to
the true model, then quantile residuals approximately follow a standard normal distribution. Quantile residuals can be assessed graphically
in terms of quantile-quantile-plots and can be an effective tool for deciding between different distributional options \citep{Klein:2013}.

\section{Simulation experiment}\label{sec:sims}

In this section, we present a simulation trial designed to show some advantages in the performance of the proposed approach with respect to
two alternative specifications of structured additive distributional regression. As a baseline, we consider Gaussian regression to describe
the conditional distribution of the response, assuming that $y_{i}$'s follow a Gaussian law. Like in our case study, Beta regression is a
useful tool to describe the conditional distribution of responses that take values in a pre-specified interval such as (0, 1). Thus,
alternatively, we assume that $y_{i}$'s follow a Beta law.
With both distributional assumptions, data are simulated from two scenarios corresponding to uncorrelated and correlated covariate repeated
measurements. For the two distributional assumptions and the two scenarios, we compare three different structured additive distributional
regression model settings: (1) as a \emph{benchmark}, we consider a model based on the ``true'' covariate values (i.e. those without
measurement error), (2) a \emph{naive} model that averages upon repeated measurements and ignores ME and (3) a model implementing the
proposed \emph{ME} adjustment. We expect that results from the latter are closer to the true model while the naive approach makes distinct
errors (larger MSE, too narrow credible intervals).

\subsection{Simulation settings}\label{sec:simuset}

For 100 simulations, $n=500$ samples of the ``true'' covariate $x_i$ ($i=1,\ldots,n$) are generated from $N(0,5)$. Then 3 replicates with
measurement error are obtained for each $x_i$ as $\tilde{\mathbf{x}}_i\thicksim N_3\left(x_i \mathbf{1}_3, \Sigma_{u;i}\right)$ with
\begin{equation}\label{eq:mecovvar}
    \Sigma_{u;i}=\sigma^2_{u;i}\left(\begin{array}{ccc} 1 & c_u & c_u \\ c_u & 1 & c_u \\ c_u & c_u & 1\end{array}\right)
\end{equation}
We allow for ME heteroscedasticity setting $\sigma^2_{u;i=1,\ldots,n/2}=1$ and $\sigma^2_{u;i=n/2+1,\ldots,n}=2$ and we consider two
alternative ME scenarios: Scenario 1 with uncorrelated replicates, i.e. $c_u=0$, and Scenario 2 where $c_u=0.8$.

For the Gaussian observation model, we simulate 500 values of the response with $y_{i}\thicksim\mbox{N}(\mu_{i},\sigma_{i}^2)$, where
$\mu_{i}=\sin(x_i)$ sets the nonlinear dependence with the covariate and we introduce response heteroscedasticity by
\[
 \mbox{Var}(y_i)=\sigma^2_{i}=\left\{\begin{array}{cc}0.5 & v_i = 1 \\0.3 & v_i = 0\end{array}\right.,\mbox{ with }v_i \sim
\mbox{Bernoulli}(0.5).
\]
We proceed in like vein for the Beta observation model and generate $n$ samples from $\mbox{Beta}(p_{i},q_{i})$ with
$\mu_{i}=p_i/(p_i+q_i)$ and $\sigma^2_i=1/(p_i+q_i+1)=\mbox{Var}(y_i)/(\mu_i(1-\mu_i))$, where $\mu_{i}=\exp(\sin(x_i))/(1+\exp(\sin(x_i))$ and
$\sigma^2_i$ is specified as in the Gaussian case.

\subsection{Model settings}

%

As for the estimation model, we consider Gaussian and Beta regression with the following settings. Assuming $y_i$'s are Gaussian with
$\mbox{N}(\mu_i,\sigma_i^2)$, model parameters are linked to regression predictors by the identity and the log link respectively:
$\mu_i=\eta^\mu(x_{i})$ and $\sigma^2_i=\exp(\eta^{\sigma^2}(x_{i}))$. Alternatively, let $y_i$'s have Beta law $\mbox{Beta}(p_{i},q_{i})$,
then
both model parameters $\mu_i$ and $\sigma^2_i$ given in Section \ref{sec:simuset} are linked to respective regression predictors
$\eta^\mu(x_{i})$ and $\eta^{\sigma^2}(x_{i})$ by the logit link, such that $\mu_i=\exp(\eta^\mu(x_{i}))/(1+\exp(\eta^\mu(x_{i})))$ and
$\sigma^2_i=\exp(\eta^{\sigma^2}(x_{i}))/(1+\exp(\eta^{\sigma^2}(x_{i})))$.

Under both distributional assumptions, we consider the two predictors as simply given by $\eta_i^\mu(x_{i})=\beta_0^\mu+f(x_{i})$ and
$\eta_i^{\sigma^2}(x_{i})=\beta_0^{\sigma^2}+\beta_1^{\sigma^2}v_i$, with $f(\cdot)$ a penalized spline term and $v_i$ defined as in
Section~\ref{sec:simuset}.

As for the measurement error, we consider the \emph{benchmark}, \emph{naive} and \emph{ME} models defined at the beginning of Section~\ref{sec:sims}. For
the latter we specify $\Sigma_{u;i}$ as in Equation \ref{eq:mecovvar}, for the two ME scenarios. Finally, we set the scaling factor $f$ in
Equation \ref{eq:propo} to $1$ in both distributional settings.

\begin{figure}
\centering
\includegraphics[scale=0.45]{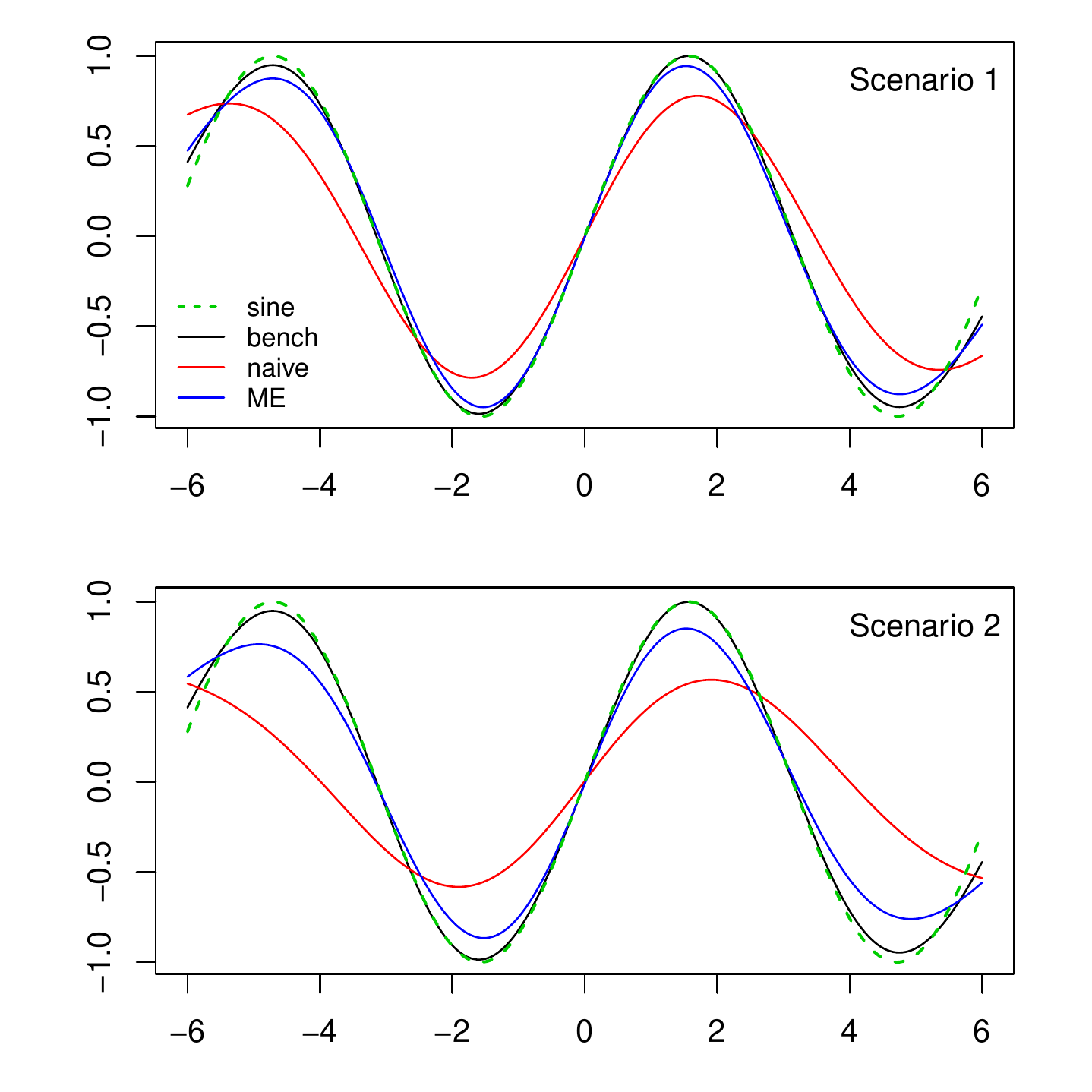}
\includegraphics[scale=0.45]{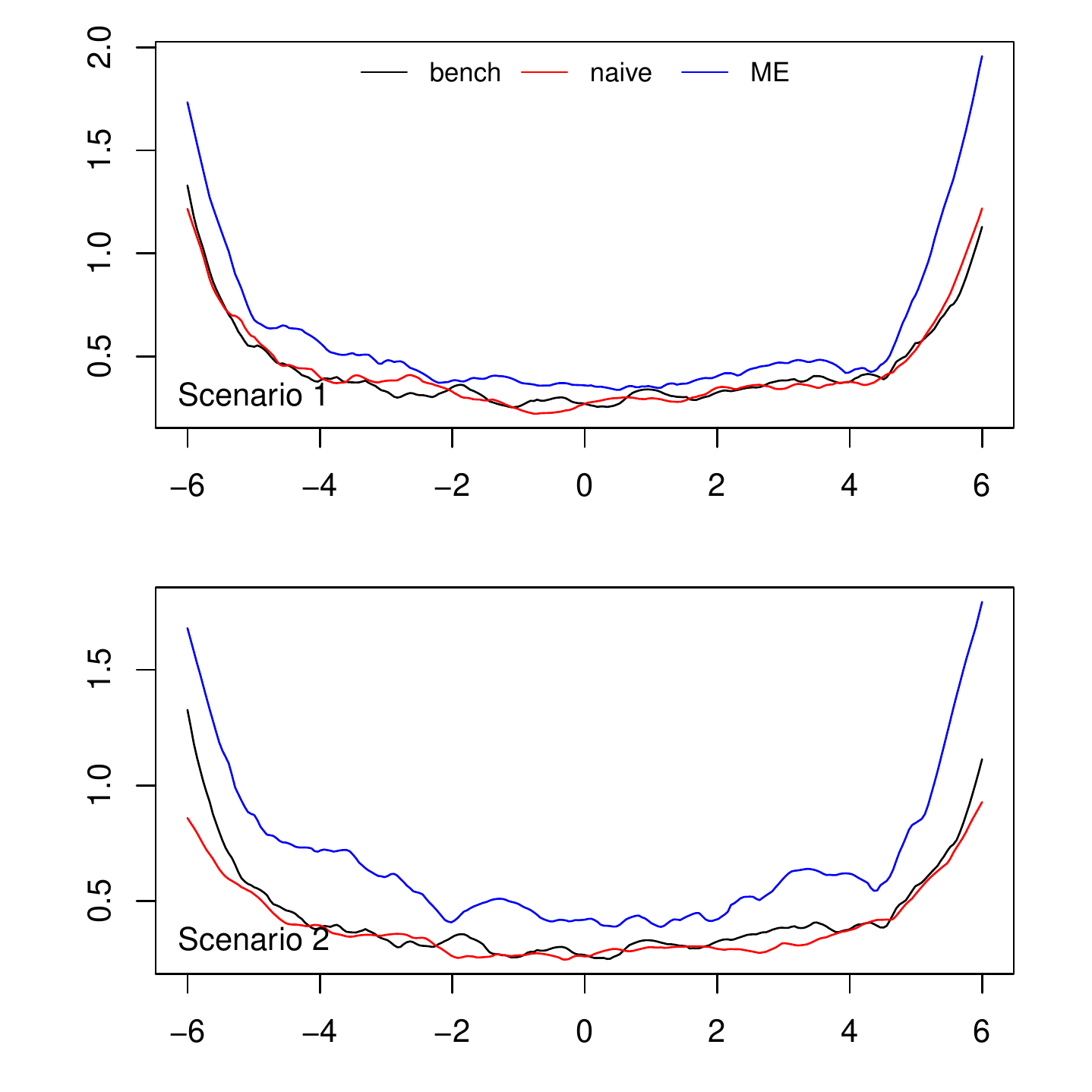}
\caption{Average estimates of the smooth effect (left) and width of 95\% credibility intervals (right) obtained with 100 simulations of the
three model settings and the two scenarios by the Gaussian distribution model.}\label{fig:smoothGauss}
\end{figure}

\begin{figure}
\centering
\includegraphics[scale=0.45]{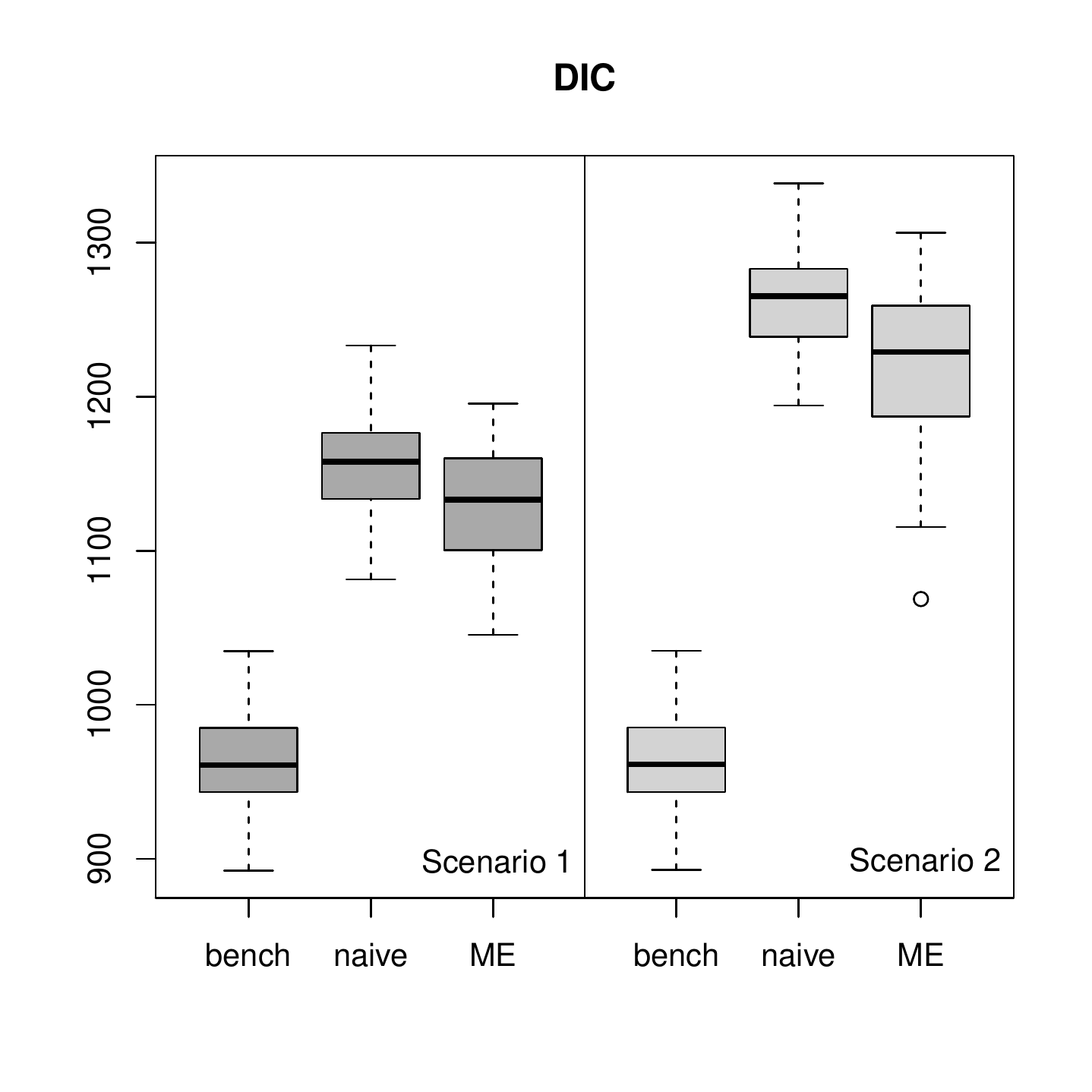}
\includegraphics[scale=0.45]{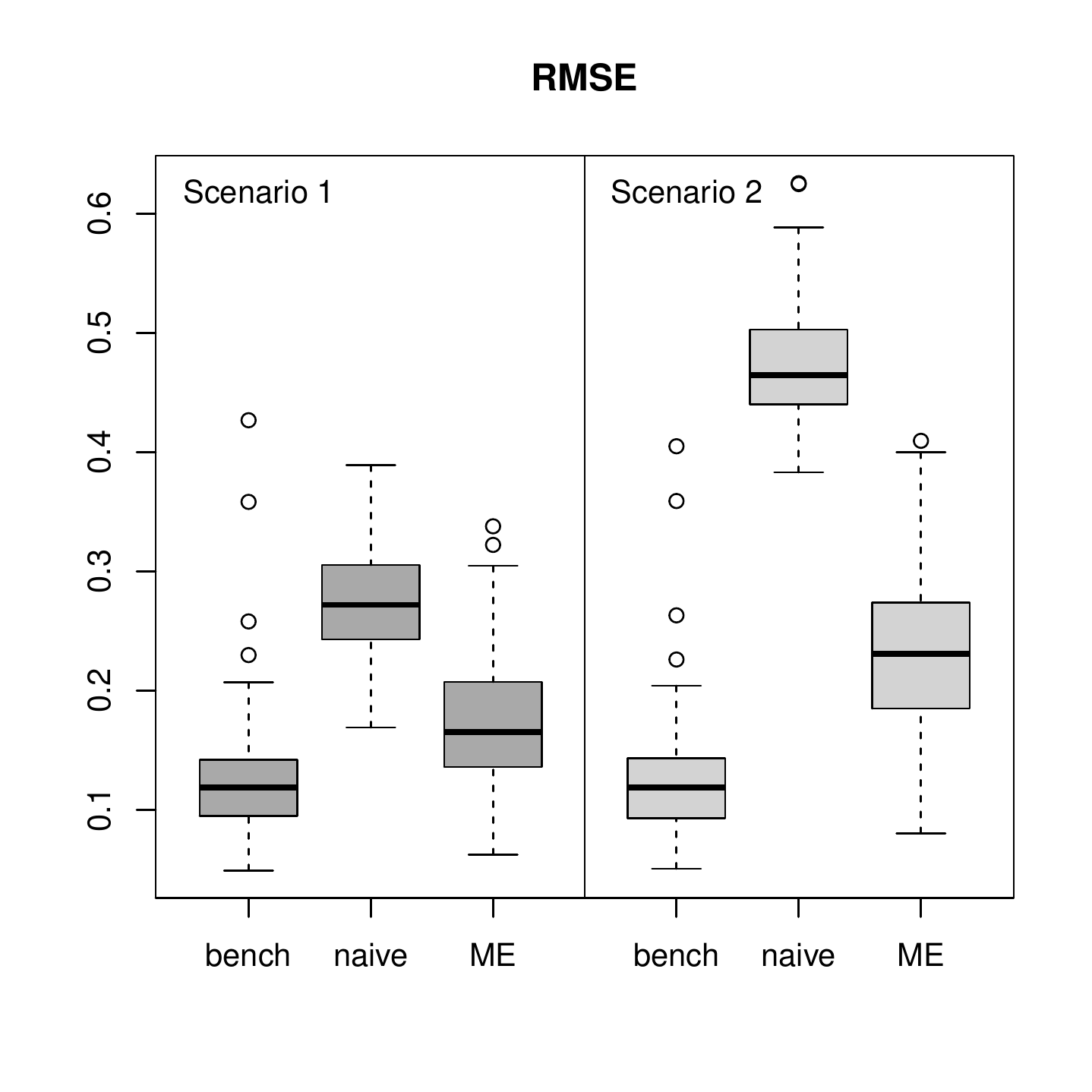}
\caption{Boxplots of DIC's (left) and RMSE's (right) of the Gaussian distribution model for 100 simulations of the three model settings and
the two scenarios.}\label{fig:boxGauss}
\end{figure}

\begin{figure}
\centering
\includegraphics[scale=0.45]{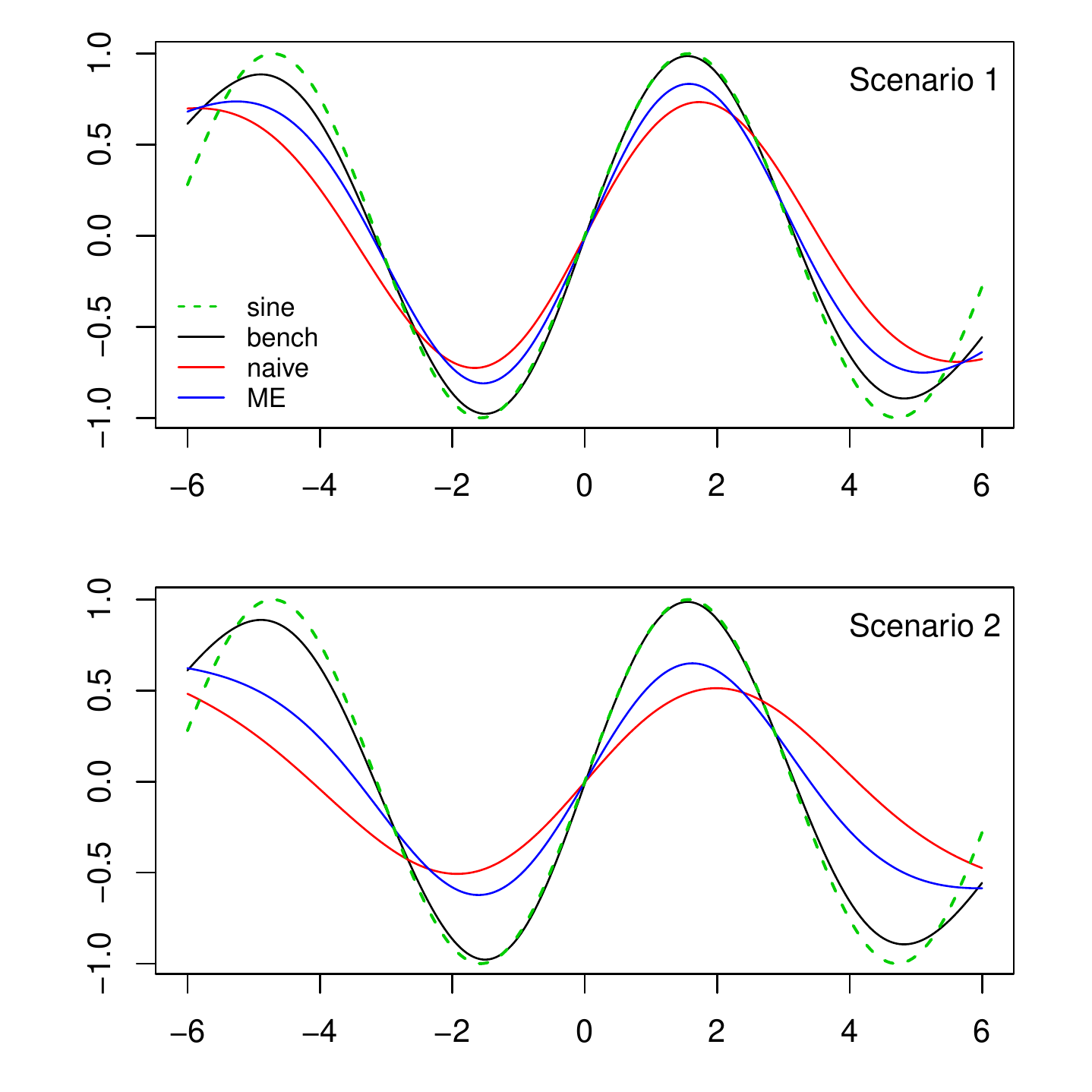}
\includegraphics[scale=0.45]{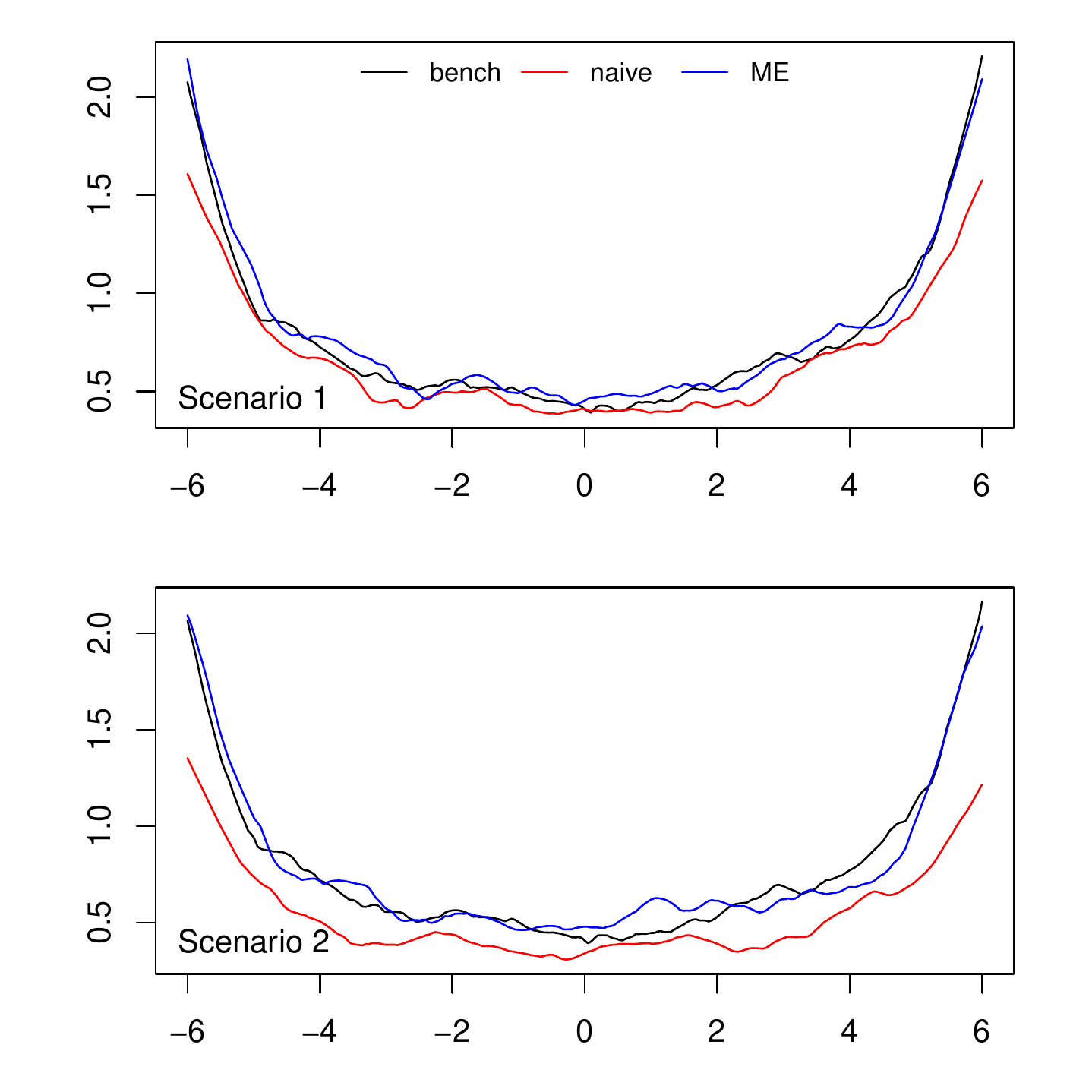}
\caption{Average estimates of the smooth effect (left) and width of 95\% credibility intervals (right) obtained with 100 simulations of the
three model settings and the two scenarios by the Beta distribution model.}\label{fig:smoothBeta}
\end{figure}

\begin{figure}
\centering
\includegraphics[scale=0.45]{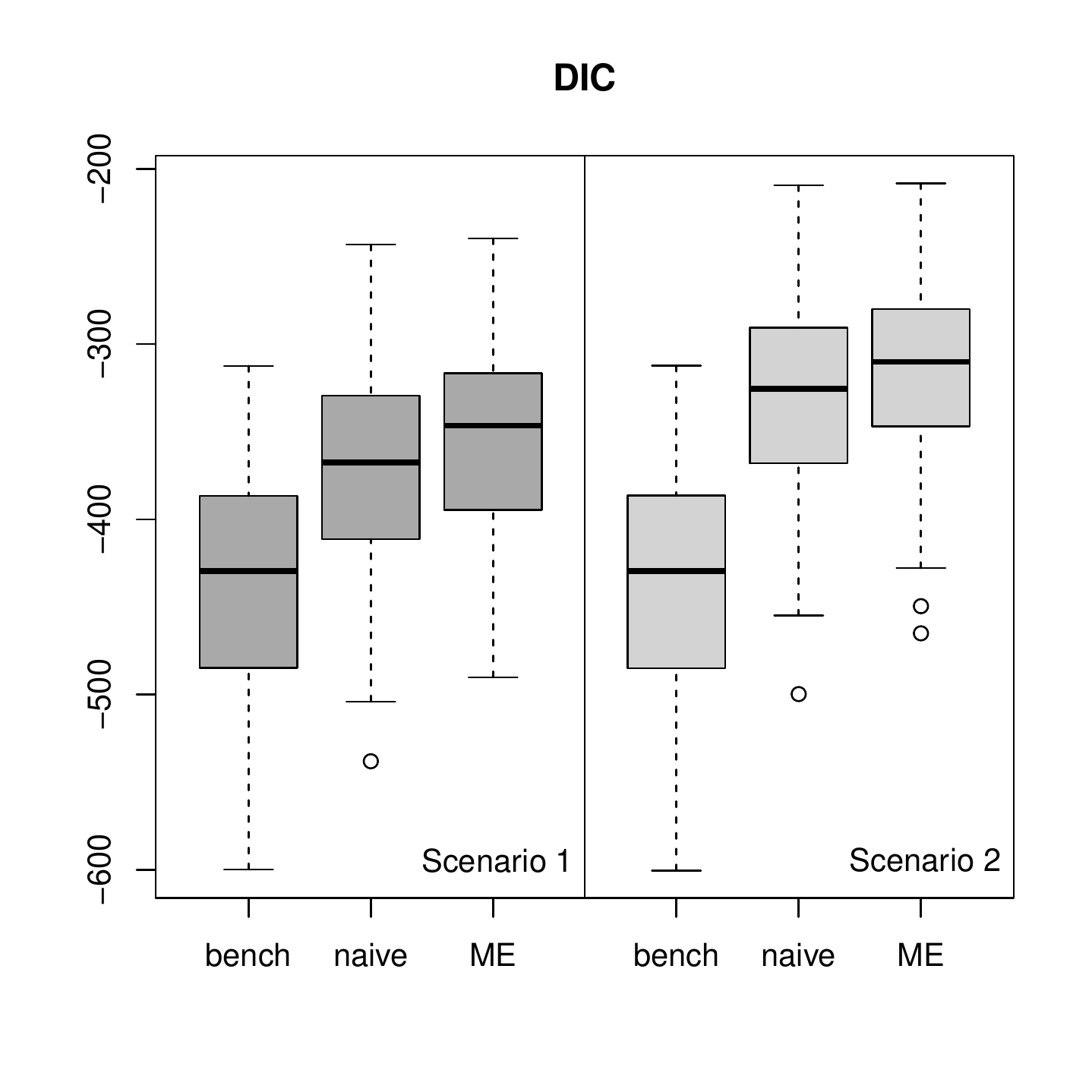}
\includegraphics[scale=0.45]{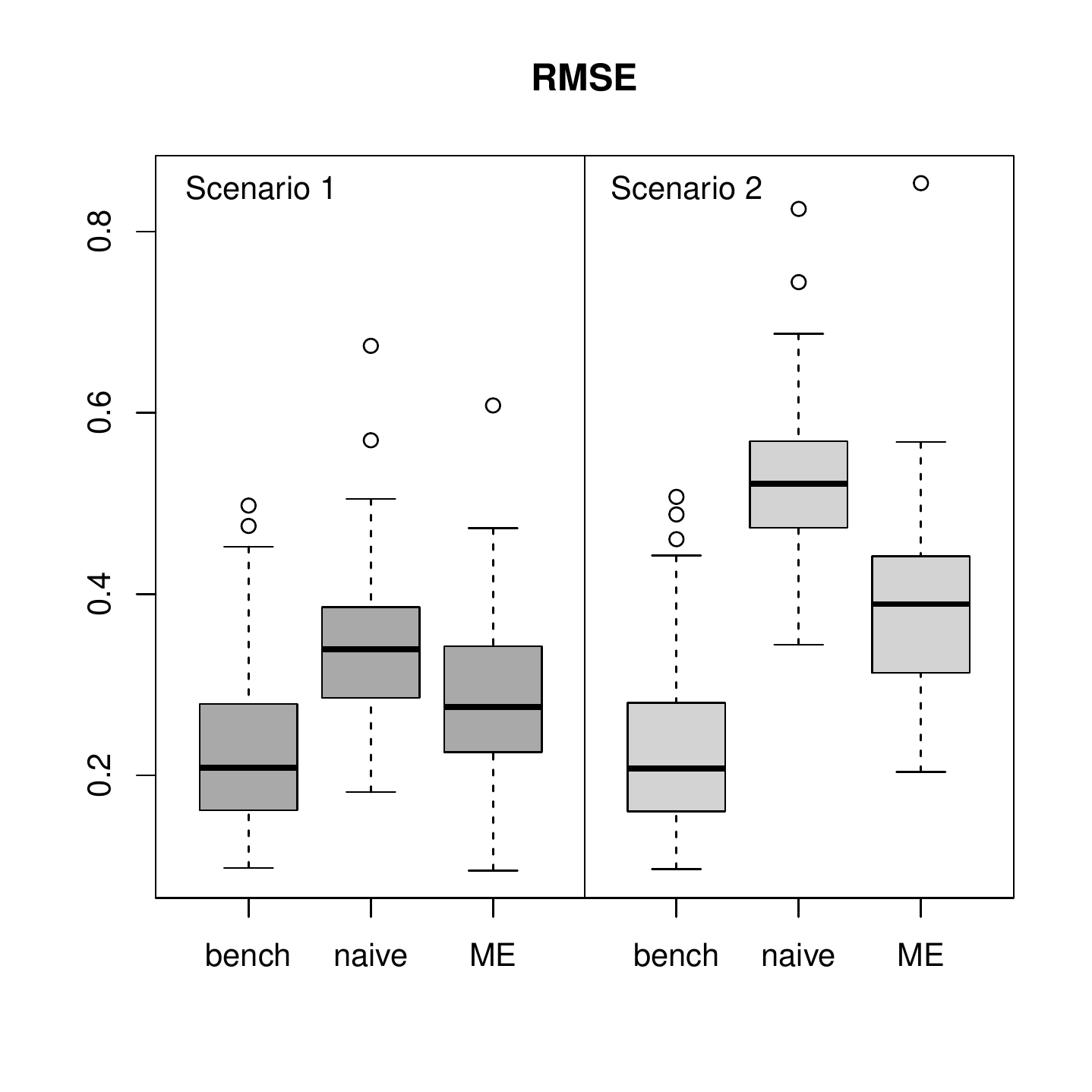}
\caption{Boxplots of DIC's (left) and RMSE's (right) of the Beta distribution model for 100 simulations of the three model settings and the
two scenarios.}\label{fig:boxBeta}
\end{figure}

\subsection{Simulation results}
In Fig. \ref{fig:smoothGauss} we show the averages and quantile ranges of smooth effect estimates obtained with 100 simulated samples for
the three Gaussian model settings and the two scenarios. While the \emph{benchmark} model setting obviously outperforms the other two, the
 \emph{ME} correction provides a sharper fit with respect to the \emph{naive} consideration of replicate averages. In general, severe
smoothing is indeed only found with the \emph{naive} model in the case of correlated measurement errors (Scenario 2). Notice that the
\emph{naive} model setting provides credible intervals as narrow as those obtained when the ME is not present (Fig. \ref{fig:smoothGauss},
right panel), thus underestimating the smooth effect variability in the presence of ME. Preference for the \emph{ME} corrected model
setting is also accorded by DIC's and RMSE's, as shown in Fig. \ref{fig:boxGauss}. The Beta distribution model obtains very similar
results, with even stronger evidence for underestimation of the smooth effect variability by the \emph{naive} model in Fig.
\ref{fig:smoothBeta} (right). With the \emph{ME} proposal, the considerable gain in terms of fit and reliability of smooth effects
variability does not fully overcome the increased complexity register by DIC's in the Beta distribution case. In this case the improvement
attained in terms of RMSE's justifies the use of this type of model compared to the \emph{naive} approach.

\section{Analysis of Sensor Data on Soil-Plant Variability}\label{sec:case}

\begin{figure}
\centering
\includegraphics[scale=0.27]{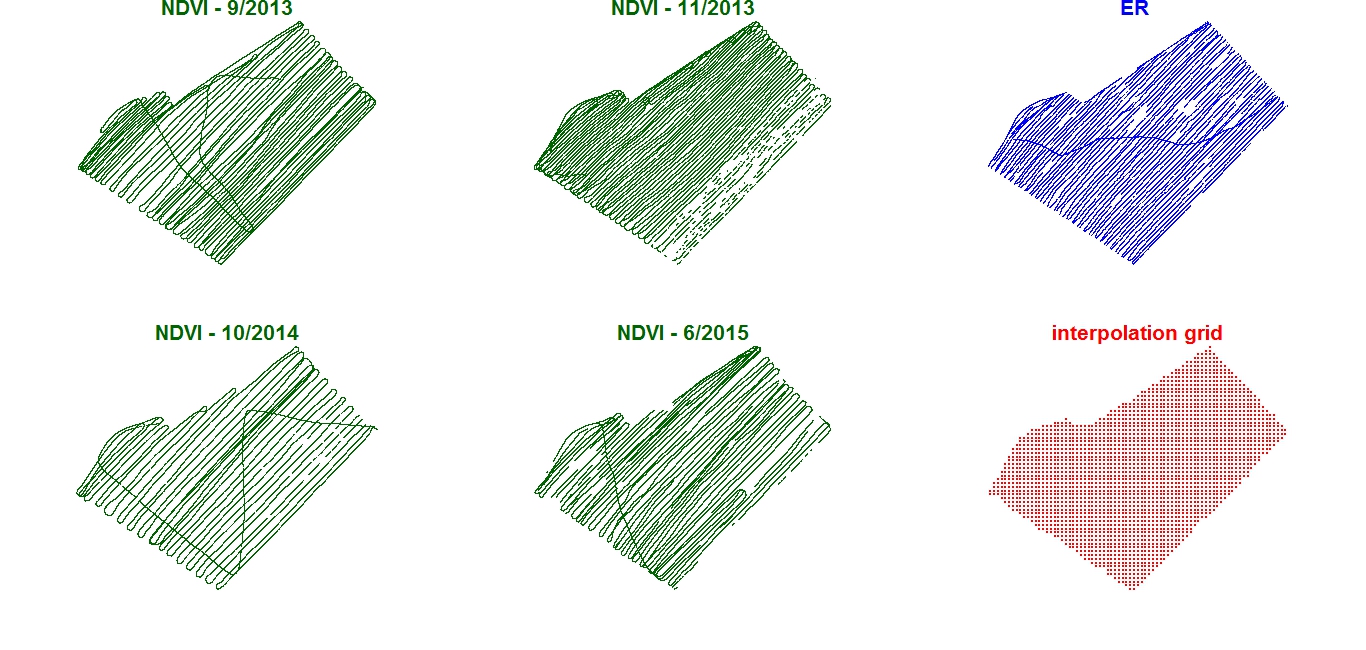}
\caption{Sampled NDVI, ER and grid point locations.}\label{fig:locat}
\end{figure}

Alfalfa (Medicago sativa L.) plays a key role in forage production systems all over the world and increasing its competitiveness is one of
the European Union's agricultural priorities \citep{visser2014}. Optimizing the use of crop inputs by applying site-specific management
strategies can be a cost-effective way to increase crop profitability and stabilize yield. To achieve this goal, precise management
activities such as precision irrigation, variable-rate N targeting, etc. are required. Precision Agriculture (PA) strategies in deep root
perennials like alfalfa require information on deep soil variability. This is due to the higher dependence of these crops, compared to
annuals, on deep water reserves often influenced by deep soil resources \citep{Dardanelli:1997}. Soil texture and structure are among the
most important soil properties affecting plant growth, since they are both strongly related to soil-nutrients and soil-water relationships
\citep{Saxton:1986}. On-the-go geophysical soil sensing has been successfully used to map the soil spatial variability at high resolution,
covering hectares in a day of work and simultaneously investigating multiple soil depths \citep{Dabas:2003, Rossi:2013}. The
non-destructive measurement of soil electrical resistivity (ER), or its inverse soil electrical conductivity, \citep{Doolittle:2014} is
correlated to many crop-relevant soil characteristics \citep{ERreview:2005}, such as the texture \citep{Banton:1997,Tetegan:2012} and
structure \citep{Besson:2004}.
We focus on the relation between plant growth and soil features as measured by two sensors within a seven hectares Alfalfa stand in
Palomonte, South Italy, with average elevation of 210 \emph{m} a.s.l.. The optical sensor GreenSeeker$^{\tiny \mbox{TM}}$ (NTech Industries
Inc., Ukiah, California, USA) measures the normalized density vegetation index (NDVI), while Automatic Resistivity Profiling
(ARP\copyright, Geocarta, Paris, F) provides multi-depth readings of the soil electrical resistivity. NDVI field measurements were taken at
four time points in different seasons, while ER measurements were taken only
once at three depth layers: 0.5\emph{m}, 1\emph{m} and 2\emph{m}. Ground-truth calibration \citep{Rossi:2015} showed that the resistitivity distribution was linked to permanent soil features such as texture, gravel content and the presence of hardpans. As shown in Table~\ref{tab:NDVIsamples} and Figure~\ref{fig:locat}, NDVI
point locations change with time and are not aligned with ER samples. Hence NDVI and ER samples are misaligned in both space and time.
Other data issues include response space-time dependence with spatially dense data \citep[big $n$ problem, ][]{JonaLasinio2013} and repeated
covariate measurements.

\begin{table}
\begin{center}
\begin{tabular}{|l||c|c|c|c||c|}
\hline & \multicolumn{4}{|c||}{NDVI} & ER \\\hline
month/year & 9/2013& 11/2013 & 10/2014 & 6/2015 & 6/2013\\
\hline
\# points &186667& 202172&91438&222278&120261\\
\hline
\end{tabular}
\end{center}
\caption{Sampling time and number of sampled locations for NDVI and ER.}\label{tab:NDVIsamples}
\end{table}

\subsection{Data pre-processing}

Big $n$ and spatial misalignment suggest to change the support of both  NDVI and ER samples.
The change of support problem \citep[COSP, see][and references therein]{gelfand2010} for continuous spatial phenomena, commonly observed at
the point and/or area level, involves a change of the spatial scale that can be required for any of several reasons, such as predicting the
process of interest at a new resolution or to fuse data coming from several sources and characterized by different spatial resolutions.
Bayesian inference with COSP may be a computationally demanding task, as it usually involves stochastic integration of the continuous
spatial process over the new support. For this reason, in case of highly complex models or huge data sets, some adjustments and model
simplifications have been proposed to make MCMC sampling feasible \citep[see][]{gelfand2010, cameletti2013}. Although relatively efficient,
these proposals don't seem to fully adapt to our setting, mostly because of the need to overcome the linear paradigm. The most commonly
used block average approach \citep[see for example][]{banerjee2014, cressie2015} would become computationally infeasible with a
semi-parametric definition of the relation between NDVI and ER.

Given the aim of this work and the data size, COSP is here addressed by a non-standard approach: the spatial resolution was downscaled by
interpolating samples to a 2574 cells square lattice overlaying the study area. Given the different number of sampled points corresponding
to each sampling occasion (NDVI) and survey (ER), we used a proportional nearest neighbors neighborhood structure to compute the downscaled
values. More precisely, 27 neighbors were just enough to obtain non-empty cells at all grid points with the least numerous NDVI series (at
the 3$^{\mbox{\textit{rd}}}$ time point). We then modified this number proportionally to the samples sizes, obtaining 55, 59, and 65
neighbors respectively for NDVI at the 1$^{\mbox{\textit{st}}}$, 2$^{\mbox{\textit{nd}}}$ and 4$^{\mbox{\textit{th}}}$ time points and 35
neighbors for ER. At each grid point we calculated the neighbors' means for both NDVI and ER, while neighbors' variances and covariances
between depth layers were obtained for ER. Summary measures of the scale and correlation of ER repeated measures at each of the 2574 grid
points provide valuable information that enables us to increase the model complexity with no additional costs in terms of parameters, i.e.
degrees of freedom (see the prior specification later in this section). Such a by-product of the downscaling of the original data is
plugged into the model likelihood.

\begin{figure}
\centering
\includegraphics[scale=0.35]{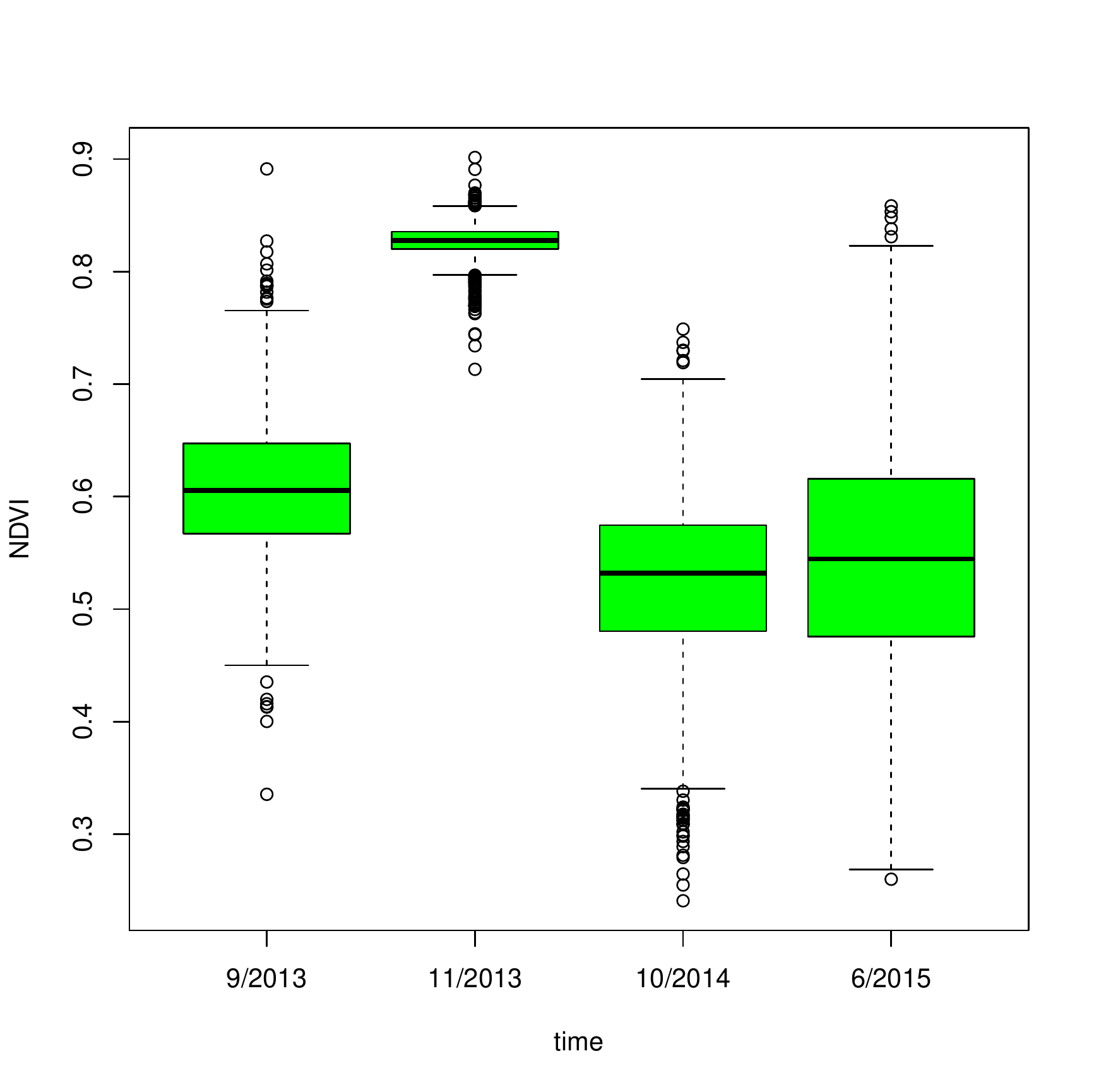}
\includegraphics[scale=0.35]{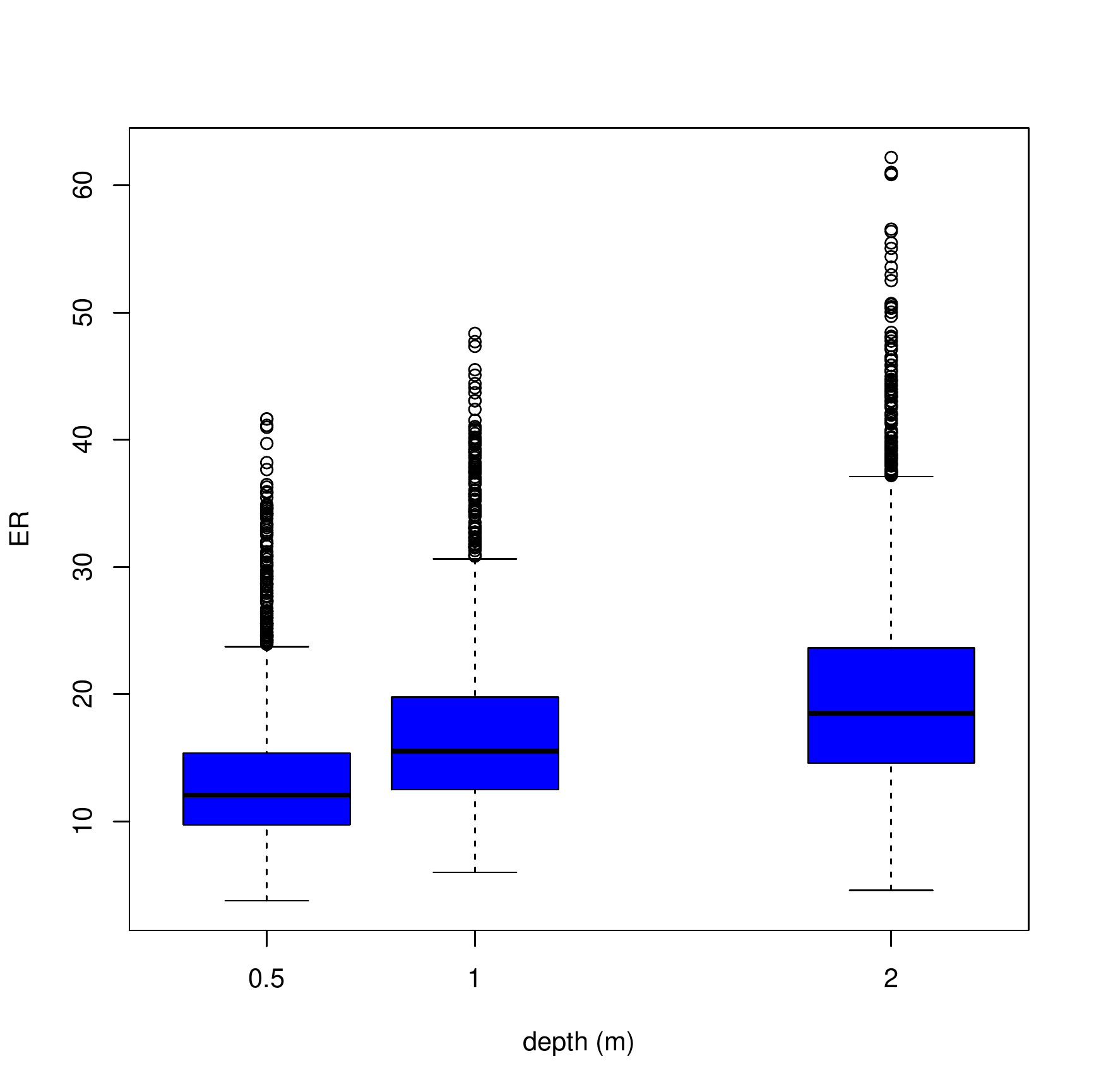}
\caption{NDVI distributions along time (left) and ER distributions along depth (right).}\label{fig:ERbox}
\end{figure}

\begin{figure}
\centering
\includegraphics[scale=0.60]{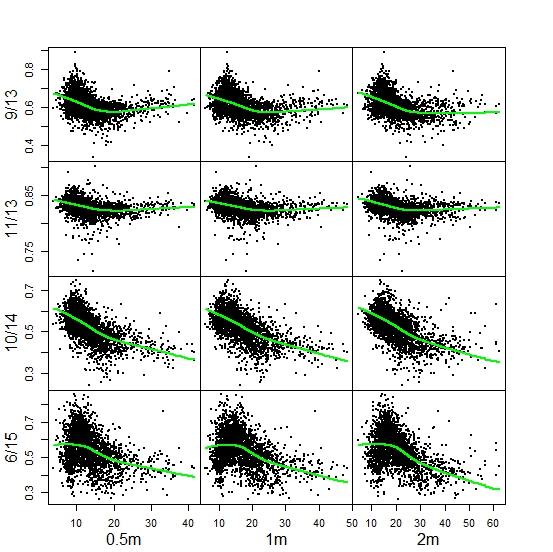}
\caption{Dependence of NDVI on ER at different time points (rows) and depth layers (columns). Lowess curves in green.}\label{fig:samples}
\end{figure}

Exploratory analysis of the interpolated data shows some interesting features that we assume to drive the specification of the
distributional regression model: NDVI has much higher position and smaller variability at the second time point, when it approached saturation (Figure~\ref{fig:ERbox},
left); ER does not show a strong systematic variation along depth (Figure~\ref{fig:ERbox}, right); the functional shape of the nonlinear
relation between NDVI and ER is common to all NDVI sampling occasions and ER depth layers (Figure~\ref{fig:samples}).

\subsection{Additive distributional regression models for location and scale}

For available NDVI recordings, we consider Gaussian and Beta distributional regression models as those defined in Section~\ref{sec:sims}
and specify the two predictors as follows. For $s=1,\ldots,2574$ grid points and $t=1,\ldots,4$ time points, the structured additive
predictor of the location parameter $\eta^\mu(\nuvec_{st})$ is determined as an additive combination of three linear and functional
effects, such as a linear seasonal effect, a spatial effect and a nonlinear effect of the continuous covariate ER:
\begin{equation}\label{eq:location}
\eta^\mu_{st}(\nuvec_{st})=\beta_0^{\mu}+\boldsymbol{i}'_t\boldsymbol{\beta}_1^{\mu}+f_1^\mu(x_s)+f_2^\mu(lon_s,lat_s)
\end{equation}
where $\boldsymbol{i}_t$ is a vector of seasonal indicator variables with cornerpoint parametrization corresponding to the first sampling
time, $f(x_s)$ is a nonlinear smooth function of latent replicate-free ER recordings $x_s$ (as in Section~\ref{subsec:me}) and $g(lon_s,
lat_s)$ is a bivariate nonlinear smooth function of geographical coordinates $lon_s$, $lat_s$.
The linear predictor of the scale parameter $\eta^{\sigma^2}(\nuvec_{st})$ is assumed to depend only on the effect of time, thus allowing heteroscedasticity of
seasonal NDVI recordings:
\begin{equation}\label{eq:scale}
\eta^{\sigma^2}(\nuvec_{st})=\beta_0^{\sigma^2}+\boldsymbol{i}'_t\boldsymbol{\beta}_1^{\sigma^2}
\end{equation}
for $t=1,\ldots,4$ time points where $\beta_0^{\sigma^2}$ and $\boldsymbol{\beta}_1^{\sigma^2}$ represent the overall level of the
predictor and the vector of seasonal effects on the transformed scale parameter.
Fixed effects $\boldsymbol{\beta}_1^{\mu}$ and $\boldsymbol{\beta}_1^{\sigma^2}$ in (\ref{eq:location}) and (\ref{eq:scale}) respectively
account for mean effects and
heteroscedasticity of NDVI seasonal recordings.
While conjugacy allowed to use Gibbs sampling to simulate from the full conditionals of the Gaussian models for the location parameter, the
Metropolis-Hastings algorithm was required to sample the posteriors of the Gaussian models for the scale parameter and those of the Beta
models.
A different simulation setup was then adopted in the two cases: for Gaussian models we obtained 10000 simulations with 5000 burnin and
thinning by 5, while Beta models required longer runs of 50000 iterations with 35000 burnin and thinning by 15. In all cases, convergence
was reached and checked by visual inspection of the trace plots and standard diagnostic tools. Fine tuning of hyperparameters lead us to 10
and 8 equidistant knots for each of the two components of the tensor product spatial smooth in the Gaussian and Beta case, respectively.

The additive distributional regression model was compared to standard additive mean regression with the same mean predictor, applying the
proposed measurement error correction to both models, under the Gaussian and Beta assumptions. By this comparison we show that adding a
structured predictor for the scale parameter improves both the in-sample and out-of-sample predictive accuracy. In the following, M1 is an
additive regression model with mean predictor as in (\ref{eq:location}), while M2 is an additive distributional regression model with the
same mean predictor and scale predictor given by (\ref{eq:scale}).


\subsection{Results}

Model comparison shows some interesting features of the proposed alternative model specifications.
Concerning the distributional assumption, DIC, WAIC and the three proper scoring rules clearly favor the Beta models (Tab. \ref{tabFit}),
showing a better compliance with in-sample and out-of-sample predictive accuracy.
As far as additive distributional regression is concerned, information criteria and scoring rules agree in assessing the proposed model (M2) as performing
better than a simple additive mean regression (M1). 

\begin{table}[ht]
\begin{center}
\begin{tabular}{|l|ccccc|}
\hline Model & DIC & WAIC & LS & SS & QS\\\hline
GM1 & -21333.3 & -21331.2 & 1.0340 & 1.8427 &3.3947\\
GM2 & -27088.4 & -27076.2& 1.1234 & 1.8986 &3.6935\\
\hline
BM1 & -22841.0 & -22835.9 & 1.1085 & 1.9298 & 3.7893\\
BM2 & -27131.4& -27126.4& 1.2140& 2.0182 & 4.2902\\\hline
\end{tabular}
\caption{Model fit statistics for Gaussian (G) and Beta (B) distribution additive mean (M1) and distributional (M2) regression models:
deviance information criterion, Watanabe-Akaike information criterion, logarithmic score, spherical score and quadratic score.
}\label{tabFit}
\end{center}
\end{table}

\begin{figure}
\centering
\includegraphics[scale=0.38]{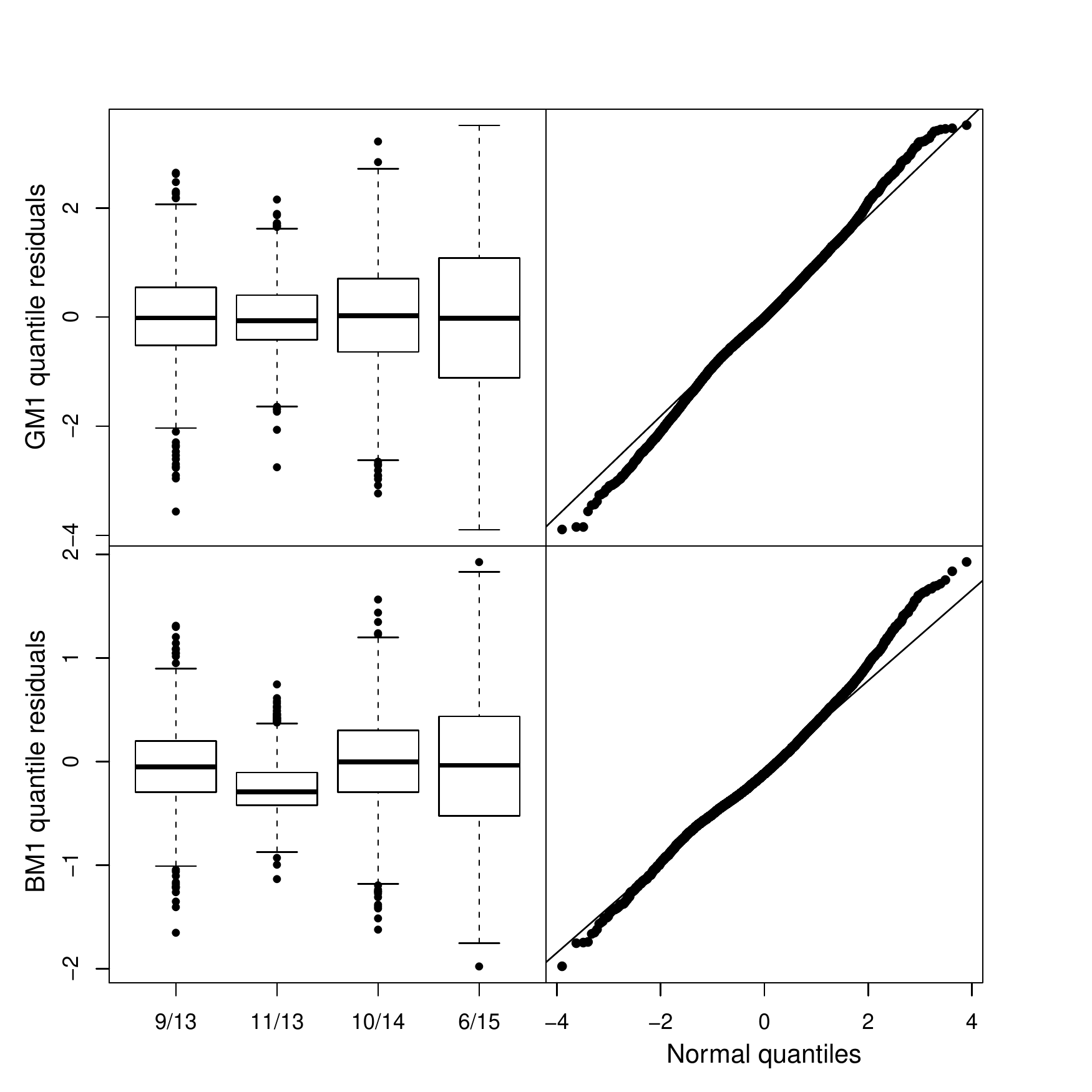}
\includegraphics[scale=0.38]{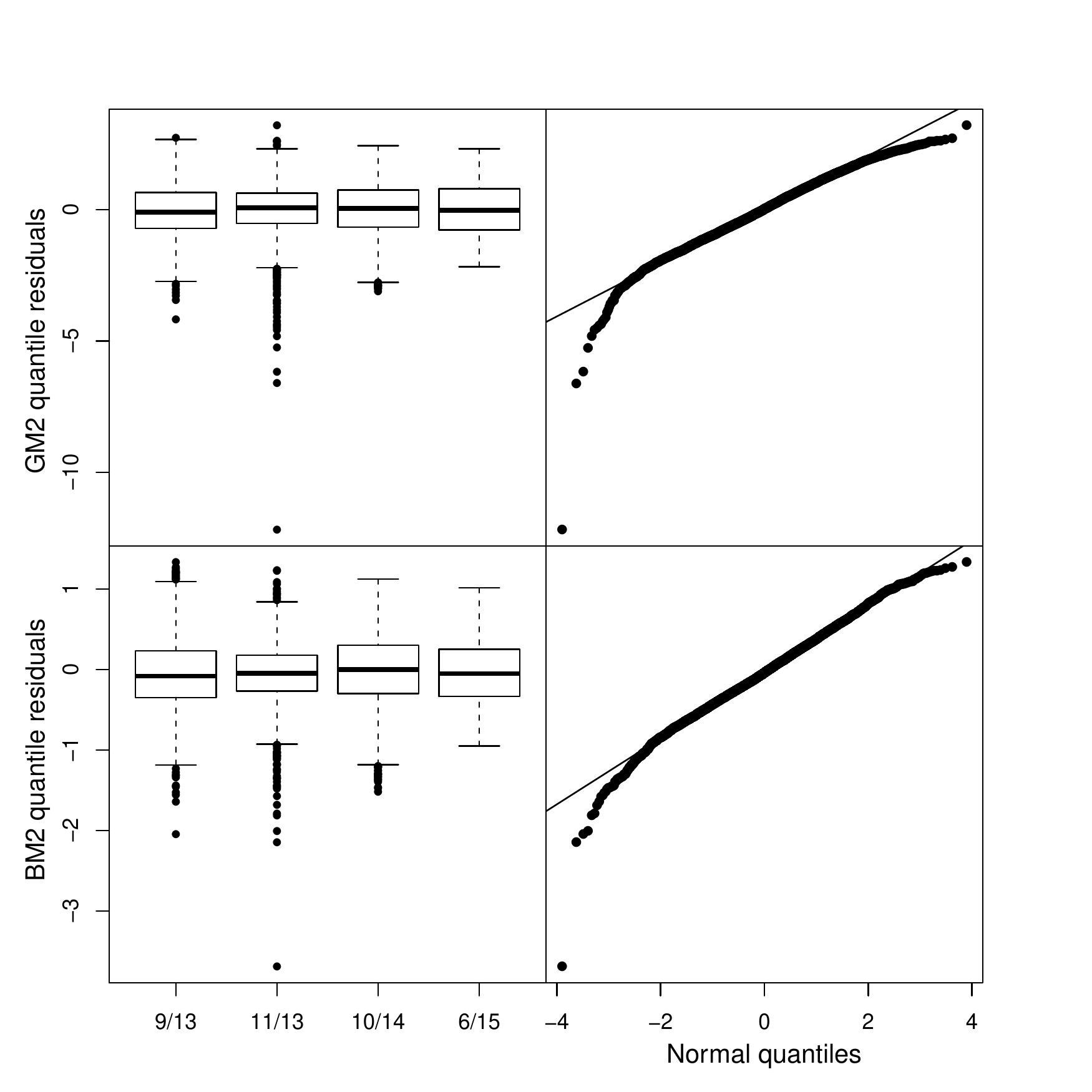}
\caption{Quantile residuals for models M1 (left) and M2 (right) under the Gaussian (top panel) and Beta (bottom panel) assumptions:
boxplots at different time points (left panel) and Normal q-q plots (right panel).}\label{fig:qres}
\end{figure}

Quantile residuals of model M2 under the two distributional assumptions show a generally good behavior, with a substantial reduction in
scale in the Beta case and only a slightly better compliance with the latter distributional shape (Figure~\ref{fig:qres}). When comparing
the two distributional assumptions, it should be recalled that Gaussian models are by far much more convenient from the computational point
of view.

Values of the fixed time effects estimates of the mean (\ref{eq:location}) and variance (\ref{eq:scale}) predictors and their 95\%
credibility intervals for the two models (Tab. \ref{tabFixed}) are expressed in the scale of the linear predictor and cannot be compared,
due to different link functions being implied. However, in both the Gaussian and Beta case their relative variation clearly reproduces the
NDVI behavior in Figure~\ref{fig:ERbox}, left.

\begin{table}[ht]
\begin{center}
{\footnotesize \begin{tabular}{|c|c|cccc|} \hline Model & Parameter & 9/13 & 11/13 & 10/14 & 6/15 \\\hline
\multirow{4}{*}{GM2} & \multirow{2}{*}{$\mu$} & 0.6037 & 0.8224 & 0.5215 & 0.5410 \\
&& {\tiny 0.5964, 0.6113} & {\tiny 0.8157, 0.8295 }& {\tiny  0.5127 0.5294} & {\tiny 0.5318 0.5503}\\
& \multirow{2}{*}{$\log(\sigma^2)$} &-5.0995 & -8.3773 & -4.5524 & -3.8718 \\
&& {\tiny -5.1588,-5.0401} & {\tiny -8.4321, -8.3190} & {\tiny -4.6071, -4.4992} & {\tiny -3.9287 -3.8171}\\\hline
\multirow{4}{*}{BM2} & \multirow{2}{*}{$\mbox{logit}(\mu)$} &0.3943 & 1.5231 & 0.0528 & 0.1365 \\
&& {\tiny 0.3528, 0.4329} & {\tiny  1.4825, 1.5636} & {\tiny0.0104 , 0.0940} & {\tiny 0.0919, 0.1799}\\
& \multirow{2}{*}{$\mbox{logit}(\sigma^2)$} &-3.7627 & -6.2897 & -3.2067 & -2.4696 \\
&& {\tiny 3.8204 ,-3.7062} & {\tiny -6.3542 ,-6.2249} & {\tiny -3.2591, -3.1520} & {\tiny -2.5245, -2.4175}\\\hline
\end{tabular}}
\caption{Fixed time effects estimates of mean and variance predictors with 95\% credibility intervals. To facilitate interpretation,
estimates were transformed avoiding the cornerpoint parametrization.}\label{tabFixed}
\end{center}
\end{table}

\begin{figure}
\centering
\includegraphics[scale=0.55]{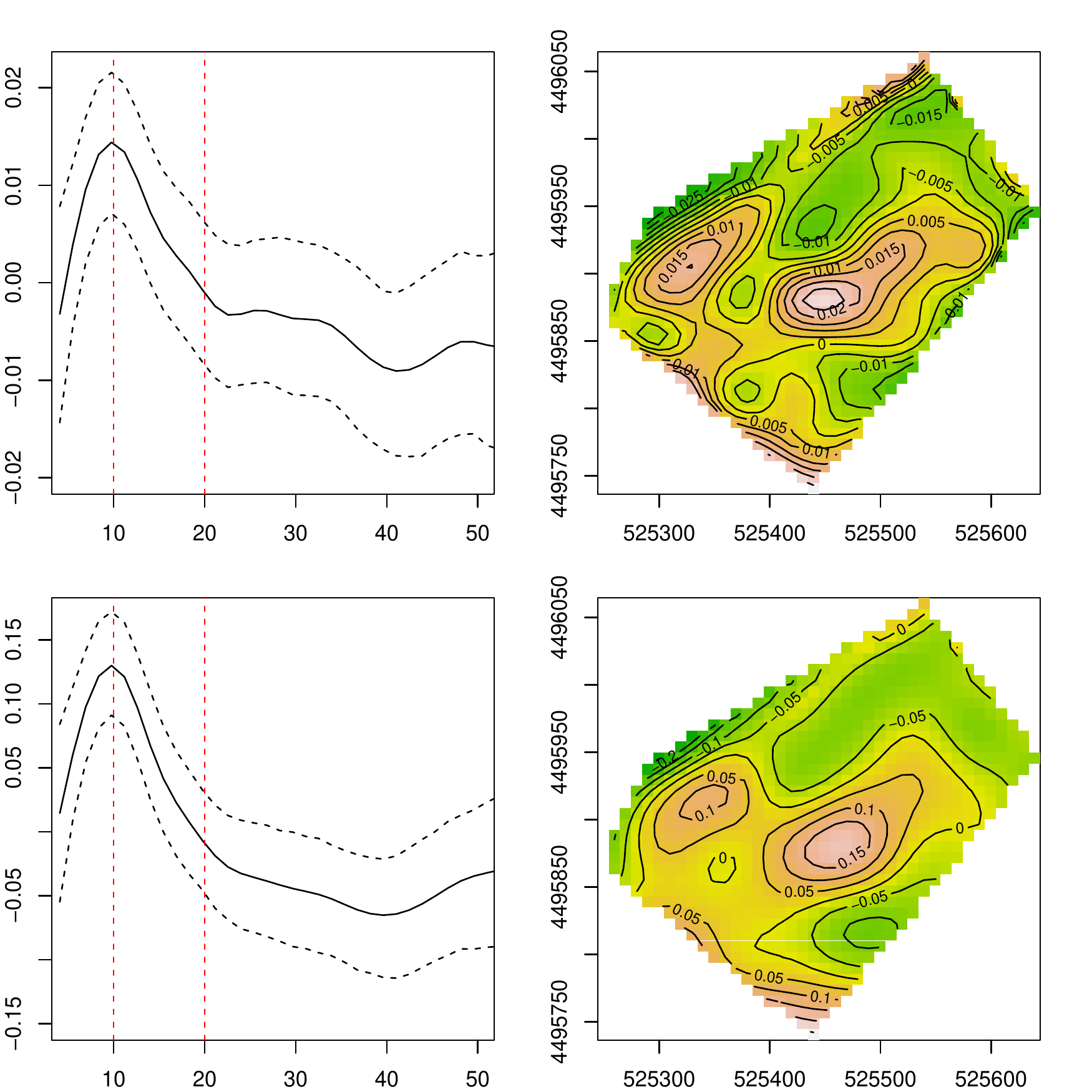}
\caption{Smooth estimates of ER effects (left) and residual spatial effects (right) for models GM2 (top), and BM2 (bottom). Notice that
while GM2 effects are on the scale of the response (NDVI), BM2 effects are estimated on the logit scale. Dotted red vertical lines locate
ER cut-offs corresponding to different monotonic soil-plant relationships.}\label{fig:smooth}
\end{figure}

Estimates of smooth effects of ER and of the nonlinear trend surface (Figure~\ref{fig:smooth}) have again different scale and common shapes
for the two distributional assumptions.
The inclusion of the nonlinear trend surface (Figure~\ref{fig:smooth}, right), besides accounting for the spatial pattern and lack of
independence between nearby observations, allows to separate the effect of ER from any other source of NDVI spatial variability including
erratic and deterministic components (such as slope and/or elevation). The estimated nonlinear effect of ER on NDVI shows a monotonically
increasing relation up to approximately 10 \emph{Ohm m} with a subsequent steep decline up to approximately 20 \emph{Ohm m}. After dropping to lower
values, the smooth function declines more slowly.
Based on the resulting estimated smooth functions (Figure~\ref{fig:smooth}, left), two ER cut-offs (at 10 and 20 \emph{Ohm m}) are proposed that can
be used to split the field in three areas characterized by a different monotonic soil-plant relationship:
\begin{itemize}
\item \textbf{Zone i: ER $<$ 10 \emph{Ohm m}}, where NDVI grows with ER and very low ER readings correspond to intermediate to high NDVI values (the former correspond  to the
presence of poorly drained soils and the consequent risk of waterlogging; crop management needs to take into account in-season rain
patterns to minimize the risks of waterlogging damages in wet years);
\item \textbf{Zone ii: 10 \emph{Ohm m} $<$ ER $<$ 20 \emph{Ohm m}}, where ER is negatively related to NDVI and soil factors affecting ER act
almost linearly and consistently on plant performance (precision management can be applied as a function of ER, i.e. the resistivity map
itself can be used as a prescription map in the corresponding areas);
\item \textbf{Zone iii: ER $>$ 20 \emph{Ohm m}}, where despite the large variation in ER there is a limited NDVI-soil responsiveness and NDVI is constantly low (corresponds to the presence of the hardpans and management criteria should differ accordingly).
\end{itemize}

Each zone conveys information on the shape and strength of the association between soil and crop variability, thus the proposed field
zonation helps discerning areas where even a little change in soil properties can affect plant productivity (zone ii) from areas where soil
environment is not practically alterable (zone iii) or in-season evaluations are possibly needed (zone i).

\section{Concluding remarks and directions of future work}\label{sec:summ}
The work described in this paper was motivated by the analysis of a database characterized by some complex features: response space-time
dependence with spatially dense data, data misalignment in both space and time and repeated covariate measurements. These data features
were addressed by first changing the spatial support of the data, then proposing an extension of structured additive distributional
regression models by the introduction of a replicated covariate measured with error. Within a fully Bayesian implementation, measurement
error is dealt with in the context of the functional modeling approach, accounting for possibly heteroscedastic and correlated covariate
replicates.
In the paper we only allow for Gaussian and Beta distributed responses, but the proposed correction is implemented to accommodate for
potentially any K-parametric family of response distributions. 
With a simulation experiment we show some advantages in the performance of the proposed ME correction with respect to two alternative less
ambitious ME specifications.
The proposed extension of the of the ME correction in \cite{Kneib2010} to structured additive regression models proves to be essential for
the case study on soil-plant sensor data, where both mean and variability effects have to be modeled. Indeed in this case the proposed
approach outperforms the simpler use of the ME correction with a mean regression model with the same (mean) predictor.

In the Bayesian framework, a straightforward extension would be the consideration of other types of (potentially non-normal) measurement
error structures, as in \citet{sarkar2014} under a structural ME approach. Given that a fully specified measurement error is given, this
will only lead to a minor adaptation of the acceptance probability in our MCMC algorithm. A more demanding extension would be to include
inference on the unknown parameters in the measurement error model, such as the covariance structure in our approach based on multivariate
normal measurement error. Such parameters will typically be hard to identify empirically unless the number of replicates and/or the sample
size is large. In the case of big spatial data, the computational burden induced by spatial correlation could be reduced integrating the ME
correction with a low rank approach \citep[see for instance][and references therein]{banerjee2014, datta2016}. While in this paper we have
considered measurement error in a covariate that enters the predictor of interest via a univariate penalized spline, other situations are
also easily conceivable. One option would be to develop a Bayesian alternative to the simulation and extrapolation algorithm developed in
\citet{KueMwaLes2006} to correct for misclassification in discrete covariates. Another route could be the consideration of measurement
error in one or both covariates entering an interaction surface modeled as a bivariate tensor product spline.\\\\

\textbf{Acknowledgments}: Alessio Pollice and Giovanna Jona Lasinio were partially supported by the PRIN2015 project "Environmental
processes and human activities: capturing their interactions via statistical methods (EPHASTAT)" funded by MIUR - Italian Ministry of
University and Research.

\bibliographystyle{natbib}
\bibliography{sensorbiblio}

\begin{thebibliography}{}

\bibitem[Arima {\em et~al.}(2017)Arima, Bell, Datta, Franco, and
  Liseo]{arima2017}
Arima, S., Bell, W.~R., Datta, G.~S., Franco, C., and Liseo, B. (2017).
\newblock Multivariate fay--herriot bayesian estimation of small area means
  under functional measurement error.
\newblock {\em Journal of the Royal Statistical Society: Series A (Statistics
  in Society)\/}, {\bf 180}(4), 1191--1209.

\bibitem[Banerjee {\em et~al.}(2014)Banerjee, Gelfand, and
  Carlin]{banerjee2014}
Banerjee, S., Gelfand, A.~E., and Carlin, B.~P. (2014).
\newblock {\em Hierarchical Modeling and Analysis for Spatial Data\/}.
\newblock Chapman and Hall/CRC, New York, second edition.

\bibitem[Banton {\em et~al.}(1997)Banton, Cimon, and Seguin]{Banton:1997}
Banton, O., Cimon, M.~A., and Seguin, M.~K. (1997).
\newblock Mapping field-scale physical properties of soil with electrical
  resistivity.
\newblock {\em Soil Science Society of America\/}, {\bf 61}(4), 1010--1017.

\bibitem[Belitz {\em et~al.}(2015)Belitz, Brezger, Kneib, Lang, and
  Umlauf]{Belitz:2013}
Belitz, C., Brezger, A., Kneib, T., Lang, S., and Umlauf, N. (2015).
\newblock {\em {BayesX}: Software for {B}ayesian Inference in Structured
  Additive Regression Models\/}.
\newblock Version 3.0.2.

\bibitem[Berry {\em et~al.}(2002)Berry, Carroll, and Ruppert]{berry2002}
Berry, S.~M., Carroll, R.~J., and Ruppert, D. (2002).
\newblock Bayesian smoothing and regression splines for measurement error
  problems.
\newblock {\em Journal of the American Statistical Association\/}, {\bf
  97}(457), 160--169.

\bibitem[Besson {\em et~al.}(2004)Besson, Cousin, Samou\"{e}lian, Boizard, and
  Richard]{Besson:2004}
Besson, A., Cousin, I., Samou\"{e}lian, A., Boizard, H., and Richard, G.
  (2004).
\newblock Structural heterogeneity of the soil tilled layer as characterized by
  2d electrical resistivity surveying.
\newblock {\em Soil and Tillage Research\/}, {\bf 79}(2), 239 -- 249.
\newblock Soil Physical Quality.

\bibitem[Brezger and Lang(2006)Brezger and Lang]{BreLan2006}
Brezger, A. and Lang, S. (2006).
\newblock Generalized structured additive regression based on {B}ayesian
  {P}-splines.
\newblock {\em Computational Statistics \& Data Analysis\/}, {\bf 50},
  967--991.

\bibitem[Cameletti(2013)Cameletti]{cameletti2013}
Cameletti, M. (2013).
\newblock The change of support problem through the inla approach.
\newblock {\em Statistica e Applicazioni, Special Issue\/}, pages 29--43.

\bibitem[Carroll {\em et~al.}(2006)Carroll, Ruppert, Stefanski, and
  Crainiceanu]{Carroll:2006}
Carroll, R.~J., Ruppert, D., Stefanski, L.~A., and Crainiceanu, C.~M. (2006).
\newblock {\em Measurement Error in Nonlinear Models: A Modern Perspective,
  Second Edition\/}.
\newblock Chapmann And Hall, CRC PRESS.

\bibitem[Cressie(2015)Cressie]{cressie2015}
Cressie, N. A.~C. (2015).
\newblock {\em Statistics for Spatial Data\/}.
\newblock John Wiley \& Sons, Inc., revised edition.

\bibitem[Dabas and Tabbagh(2003)Dabas and Tabbagh]{Dabas:2003}
Dabas, M. and Tabbagh, A. (2003).
\newblock A comparison of emi and dc methods used in soil mapping - theoretical
  considerations for precision agriculture.
\newblock In J.~Stafford and A.~Warner, editors, {\em Precision Agriculture\/},
  pages 121--127. Wageningen Academic Publishers.

\bibitem[Dardanelli {\em et~al.}(1997)Dardanelli, Bachmeier, Sereno, and
  Gil]{Dardanelli:1997}
Dardanelli, J., Bachmeier, O., Sereno, R., and Gil, R. (1997).
\newblock Rooting depth and soil water extraction patterns of different crops
  in a silty loam haplustoll.
\newblock {\em Field Crops Research\/}, {\bf 54}(1), 29 -- 38.

\bibitem[Datta {\em et~al.}(2016)Datta, Banerjee, Finley, and
  Gelfand]{datta2016}
Datta, A., Banerjee, S., Finley, A.~O., and Gelfand, A.~E. (2016).
\newblock Hierarchical nearest-neighbor gaussian process models for large
  geostatistical datasets.
\newblock {\em Journal of the American Statistical Association\/}, {\bf
  111}(514), 800--812.

\bibitem[Doolittle and Brevik(2014)Doolittle and Brevik]{Doolittle:2014}
Doolittle, J.~A. and Brevik, E.~C. (2014).
\newblock The use of electromagnetic induction techniques in soils studies.
\newblock {\em Geoderma\/}, {\bf 223--225}, 33 -- 45.

\bibitem[Dunn and Smyth(1996)Dunn and Smyth]{Dunn:1996}
Dunn, P.~K. and Smyth, G.~K. (1996).
\newblock Randomized quantile residuals.
\newblock {\em Journal of Computational and Graphical Statistics\/}, {\bf
  5}(3), 236--244.

\bibitem[Fahrmeir {\em et~al.}(2013)Fahrmeir, Kneib, Lang, and
  Marx]{Fahrmeir:2013}
Fahrmeir, L., Kneib, T., Lang, S., and Marx, B. (2013).
\newblock {\em Regression Models, Methods and Applications\/}.
\newblock Springer Verlag.

\bibitem[Ferrari and Cribari-Neto(2004)Ferrari and Cribari-Neto]{FerCri2004}
Ferrari, S. L.~P. and Cribari-Neto, F. (2004).
\newblock Beta regression for modelling rates and proportions.
\newblock {\em Journal of Applied Statistics\/}, {\bf 31}, 799--815.

\bibitem[Gelfand {\em et~al.}(2010)Gelfand, Diggle, Guttorp, and
  Fuentes]{gelfand2010}
Gelfand, A.~E., Diggle, P., Guttorp, P., and Fuentes, M. (2010).
\newblock {\em Handbook of Spatial Statistics\/}.
\newblock Chapman \& Hall-CRC Handbooks of Modern Statistical Methods. Chapmann
  And Hall, CRC PRESS.

\bibitem[Gelman {\em et~al.}(2014)Gelman, Hwang, and Vehtari]{Gelman2014}
Gelman, A., Hwang, J., and Vehtari, A. (2014).
\newblock Understanding predictive information criteria for bayesian models.
\newblock {\em Statistics and Computing\/}, {\bf 24}(6), 997--1016.

\bibitem[Gneiting and Raftery(2007)Gneiting and Raftery]{Gneiting:2007}
Gneiting, T. and Raftery, A.~E. (2007).
\newblock Strictly proper scoring rules, prediction, and estimation.
\newblock {\em Journal of the American Statistical Association\/}, {\bf
  102}(477), 359--378.

\bibitem[Huque {\em et~al.}(2016)Huque, Bondell, Carroll, and Ryan]{huque2016}
Huque, M.~H., Bondell, H.~D., Carroll, R.~J., and Ryan, L.~M. (2016).
\newblock Spatial regression with covariate measurement error: A semiparametric
  approach.
\newblock {\em Biometrics\/}, {\bf 72}(3), 678--686.

\bibitem[Jona~Lasinio {\em et~al.}(2013)Jona~Lasinio, Mastrantonio, and
  Pollice]{JonaLasinio2013}
Jona~Lasinio, G., Mastrantonio, G., and Pollice, A. (2013).
\newblock Discussing the ``big n problem''.
\newblock {\em Statistical Methods {\&} Applications\/}, {\bf 22}(1), 97--112.

\bibitem[Klein {\em et~al.}(2013)Klein, Kneib, and Lang]{Klein:2013}
Klein, N., Kneib, T., and Lang, S. (2013).
\newblock Bayesian structured additive distributional regression.
\newblock Working Papers in Economics and Statistics 2013-23, University of
  Innsbruck.

\bibitem[Klein {\em et~al.}(2015a)Klein, Kneib, Klasen, and Lang]{Klein:2015}
Klein, N., Kneib, T., Klasen, S., and Lang, S. (2015a).
\newblock Bayesian structured additive distributional regression for
  multivariate responses.
\newblock {\em Journal of the Royal Statistical Society: Series C (Applied
  Statistics)\/}, {\bf 64}(4), 569--591.

\bibitem[Klein {\em et~al.}(2015b)Klein, Kneib, Lang, and
  Sohn]{KleSohKneLan2015}
Klein, N., Kneib, T., Lang, S., and Sohn, A. (2015b).
\newblock {B}ayesian structured additive distributional regression with with an
  application to regional income inequality in germany.
\newblock {\em Annals of Applied Statistics\/}, {\bf 9}, 1024---1052.

\bibitem[Kneib {\em et~al.}(2010)Kneib, Brezger, and Crainiceanu]{Kneib2010}
Kneib, T., Brezger, A., and Crainiceanu, C.~M. (2010).
\newblock Generalized semiparametric regression with covariates measured with
  error.
\newblock In T.~Kneib and G.~Tutz, editors, {\em Statistical Modelling and
  Regression Structures: Festschrift in Honour of Ludwig Fahrmeir\/}, pages
  133--154. Physica-Verlag HD, Heidelberg.

\bibitem[Kneib {\em et~al.}(2017)Kneib, Klein, Lang, and
  Umlauf]{KneKleLanUml2017}
Kneib, T., Klein, N., Lang, S., and Umlauf, N. (2017).
\newblock Modular regression -- a lego system for building structured additive
  distributional regression models with tensor product interactions.
\newblock Technical report.

\bibitem[K{\"u}chenhoff {\em et~al.}(2006)K{\"u}chenhoff, Mwalili, and
  Lesaffre]{KueMwaLes2006}
K{\"u}chenhoff, H., Mwalili, S.~M., and Lesaffre, E. (2006).
\newblock A general method for dealing with misclassification in regression:
  the misclassification simex.
\newblock {\em Biometrics\/}, {\bf 62}, 85--96.

\bibitem[Lang {\em et~al.}(2014a)Lang, Umlauf, Wechselberger, Harttgen, and
  Kneib]{Lang:2014}
Lang, S., Umlauf, N., Wechselberger, P., Harttgen, K., and Kneib, T. (2014a).
\newblock Multilevel structured additive regression.
\newblock {\em Statistics and Computing\/}, {\bf 24}(2), 223--238.

\bibitem[Lang {\em et~al.}(2014b)Lang, Umlauf, Wechselberger, Harttgen, and
  Kneib]{LanUmlWecHarKne2012}
Lang, S., Umlauf, N., Wechselberger, P., Harttgen, K., and Kneib, T. (2014b).
\newblock Multilevel structured additive regression.
\newblock {\em Statistics and Computing\/}, {\bf 24}, 223--238.

\bibitem[Loken and Gelman(2017)Loken and Gelman]{Loken584}
Loken, E. and Gelman, A. (2017).
\newblock Measurement error and the replication crisis.
\newblock {\em Science\/}, {\bf 355}(6325), 584--585.

\bibitem[Muff {\em et~al.}(2015)Muff, Riebler, Held, Rue, and Saner]{muff2015}
Muff, S., Riebler, A., Held, L., Rue, H., and Saner, P. (2015).
\newblock Bayesian analysis of measurement error models using integrated nested
  laplace approximations.
\newblock {\em Journal of the Royal Statistical Society: Series C (Applied
  Statistics)\/}, {\bf 64}(2), 231--252.

\bibitem[Rossi {\em et~al.}(2013)Rossi, Pollice, Diago, Oliveira, Millan,
  Bitella, Amato, and Tardaguila]{Rossi:2013}
Rossi, R., Pollice, A., Diago, M.-P., Oliveira, M., Millan, B., Bitella, G.,
  Amato, M., and Tardaguila, J. (2013).
\newblock Using an automatic resistivity profiler soil sensor on-the-go in
  precision viticulture.
\newblock {\em Sensors (Basel)\/}, {\bf 13}(1), 1121--1136.

\bibitem[Rossi {\em et~al.}(2015)Rossi, Pollice, Bitella, Bochicchio,
  D'Antonio, Alromeed, Stellacci, Labella, and Amato]{Rossi:2015}
Rossi, R., Pollice, A., Bitella, G., Bochicchio, R., D'Antonio, A., Alromeed,
  A.~A., Stellacci, A.~M., Labella, R., and Amato, M. (2015).
\newblock Soil bulk electrical resistivity and forage ground cover: nonlinear
  models in an alfalfa (medicago sativa l.) case study.
\newblock {\em Italian Journal of Agronomy\/}, {\bf 10}(4), 215--219.

\bibitem[Samou\"{e}lian {\em et~al.}(2005)Samou\"{e}lian, Cousin, Tabbagh,
  Bruand, and Richard]{ERreview:2005}
Samou\"{e}lian, A., Cousin, I., Tabbagh, A., Bruand, A., and Richard, G.
  (2005).
\newblock Electrical resistivity survey in soil science: a review.
\newblock {\em Soil and Tillage Research\/}, {\bf 83}(2), 173--193.

\bibitem[Sarkar {\em et~al.}(2014)Sarkar, Mallick, and Carroll]{sarkar2014}
Sarkar, A., Mallick, B.~K., and Carroll, R.~J. (2014).
\newblock Bayesian semiparametric regression in the presence of conditionally
  heteroscedastic measurement and regression errors.
\newblock {\em Biometrics\/}, {\bf 70}(4), 823--834.

\bibitem[Saxton {\em et~al.}(1986)Saxton, Rawls, Romberger, and
  Papendick]{Saxton:1986}
Saxton, K.~E., Rawls, W., Romberger, J.~S., and Papendick, R.~I. (1986).
\newblock Estimating generalized soil-water characteristics from texture.
\newblock {\em Soil Science Society of America Journal\/}, {\bf 50}(4),
  1031--1036.

\bibitem[Schreuder and de~Visser(2014)Schreuder and de~Visser]{visser2014}
Schreuder, R. and de~Visser, C. (2014).
\newblock Report eip-agri focus group protein crops.
\newblock Technical report, European Commission.

\bibitem[Spiegelhalter {\em et~al.}(2002)Spiegelhalter, Best, Carlin, and Van
  Der~Linde]{Spiegelhalter:2002}
Spiegelhalter, D.~J., Best, N.~G., Carlin, B.~P., and Van Der~Linde, A. (2002).
\newblock Bayesian measures of model complexity and fit.
\newblock {\em Journal of the Royal Statistical Society: Series B (Statistical
  Methodology)\/}, {\bf 64}(4), 583--639.

\bibitem[Tetegan {\em et~al.}(2012)Tetegan, Pasquier, Besson, Nicoullaud,
  Bouthier, Bourennane, Desbourdes, King, and Cousin]{Tetegan:2012}
Tetegan, M., Pasquier, C., Besson, A., Nicoullaud, B., Bouthier, A.,
  Bourennane, H., Desbourdes, C., King, D., and Cousin, I. (2012).
\newblock Field-scale estimation of the volume percentage of rock fragments in
  stony soils by electrical resistivity.
\newblock {\em \{CATENA\}\/}, {\bf 92}, 67 -- 74.

\bibitem[Watanabe(2010)Watanabe]{Watanabe:2010}
Watanabe, S. (2010).
\newblock Asymptotic equivalence of bayes cross validation and widely
  applicable information criterion in singular learning theory.
\newblock {\em Journal of Machine Learning Research\/}, {\bf 11}, 3571--3594.

\end{thebibliography}

\end{document}